\documentclass[preprint]{IEEEtran}

\usepackage{cite}      
\usepackage{graphicx}  
\usepackage{algorithm,algpseudocode} 
\usepackage{algorithmicx}
\usepackage{multirow}

\usepackage{amssymb}
\usepackage{mathrsfs}
\usepackage{amsmath}

\usepackage[sans]{dsfont}
\usepackage{stmaryrd}
\usepackage{pifont}
\usepackage{wasysym}

\usepackage{array}
\usepackage{url}
\usepackage[usenames]{color}

\usepackage{epstopdf}
\usepackage{tabularx}
\usepackage{caption}
\usepackage{subcaption}

\newtheorem{innercustomthm}{Theorem}
\newenvironment{customthm}[1]
  {\renewcommand\theinnercustomthm{#1}\innercustomthm}
  {\endinnercustomthm}

\newtheorem{example}{Example}

\newtheorem{lemma}{Lemma}

\def\cC{\mathcal C}

\def\cM{\mathcal M}
\def\cN{\mathcal N}

\newcommand{\hide}[1]{}

\newcolumntype{x}[1]{>{\centering\arraybackslash\hspace{0pt}}m{#1}}

\begin{document}

\title{Online Algorithms for Information Aggregation from Distributed and Correlated Sources}

\author{Chi-Kin~Chau, {\em Member}, {\em IEEE}, Majid Khonji, and Muhammad Aftab
\thanks{C.-K. Chau, M. Khonji, and M. Aftab are with the Department of EECS at Masdar Institute of Science and Technology, UAE (e-mail: \{ckchau, mkhonji, muhaftab\}@masdar.ac.ae).}
}

\pagestyle{plain}
\thispagestyle{plain}

\maketitle

\begin{abstract}
There is a fundamental trade-off between the communication cost and latency in information aggregation. Aggregating multiple communication messages over time can alleviate overhead and improve energy efficiency on one hand, but inevitably incurs information delay on the other hand. In the presence of uncertain future inputs, this trade-off should be balanced in an online manner, which is studied by the classical {\em dynamic TCP ACK problem} for a single information source. In this paper, we extend dynamic TCP ACK problem to a general setting of collecting aggregate information from distributed and correlated information sources. In this model, distributed sources observe correlated events, whereas only a small number of reports are required from the sources. The sources make online decisions about their reporting operations in a distributed manner without prior knowledge of the local observations at others. Our problem captures a wide range of applications, such as in-situ sensing, anycast acknowledgement and distributed caching. We present simple threshold-based competitive {\em distributed online algorithms} under different settings of intercommunication. Our algorithms match the theoretical lower bounds in order of magnitude. We observe that our algorithms can produce satisfactory performance in simulations and practical testbed.

\end{abstract}

\begin{IEEEkeywords}
Online algorithms, competitive analysis, distributed optimization problem, latency, communication cost
\end{IEEEkeywords}

%\vspace{-5pt}
\section{Introduction} \label{sec:intro}

Information often arises from multiple sources, and efficient mechanisms are required to aggregate the information from these sources. One can aggregate multiple communication messages to alleviate overhead and improve energy efficiency. However, there is a fundamental trade-off between the communication cost and latency in information aggregation. Aggregating multiple communication messages over time can reduce the communication cost on one hand, but inevitably incurs information delay on the other hand. Typically, there is uncertainty in future inputs, and the balance between communication cost and latency needs to be tackled in an online manner. A well-known result is the classical dynamic TCP ACK problem \cite{Dooly:2001}, which studies the balance between communication cost (i.e., the number of ACKs) and latency (i.e., the lag between packet reception and ACK transmission) in an online manner for a single information source. 

Notably, there are similar trade-off balancing problems in a wide range of information exchange applications, especially when there are distributed systems sharing correlated local information. A critical problem is how to coordinate the operations of these systems for jointly balancing the communication cost and latency of information exchanges. There are several concrete examples as follows:
\begin{itemize}

\item {\em In-situ Sensing}:  A set of sensors are assigned to monitor certain events of interest in their vicinity based on local measurements. The measurements are correlated and are required to report to a base station. The communication messages from sensors should be optimized to reduce battery consumption and improve operation lifetime, while incurring tolerable latency of reporting.
%In particular, there is a relatively large data frame size, and considerable connection setup overhead. Also, the transceivers usually consume a large portion of energy in generating high frequency, regardless of the packet size. Therefore, reducing the number of communications can effectively alleviate the battery consumption. 

\item {\em Anycast ACKs}: A sender is transmitting a number of packets to a group of anycast receivers in an unreliable channel. An ACK from one of the successful receivers is required to confirm the reception. Furthermore, the receiver may reply an ACK to a batch of packets, instead of to every packet. To improve throughput, the number of ACKs needs to be minimized among the receivers, taking into account the incurred latency.

\item {\em Distributed Caching}: For distributed storage and web caching services, the cached objects in data servers are often represented by a hash file, which is distributed among other servers for coordinating information retrieval. The total size of hash files generated by all servers should be optimized, by taking into account the commonly cached objects and the latency of distributing these hash files.

\end{itemize}

The aforementioned applications often feature a trade-off between a certain communication cost (e.g., energy consumption, bandwidth and storage space) and the incurred latency in an online fashion, similar to that of dynamic TCP ACK problem. However, as a departure from the classical TCP ACK problem, these problems involve multiple information sources, and the difficulty is that the local information of each source is typically unknown to others without communication, which hinders the joint optimization. We consider a general problem of a general setting of collecting aggregate information from distributed information sources as follows:
\begin{itemize}

\item {\em Online Decisions:} A group of systems are tasked with reporting certain correlated events (e.g., sensing events, receptions of packets, and cached objects). We assume no prior knowledge on the arrivals of events (either stochastic or non-stochastic models). The systems are required to make decisions in an online manner.

\item {\em Distributed Decisions:} The local decisions of the systems about their report messages should globally optimize the communication cost and the incurred latency. In this paper, we consider various settings: (i) {\em no intercommunication}, where each system is not aware of the communications of others, (ii) {\em full intercommunication}, where every system can overhear the communications of others, and (iii) {\em partial intercommunication}, such that the extent of intercommunication is characterized by a communication graph. 

\end{itemize}

Not only generalizing the dynamic TCP ACK problem to a setting with distributed information sources, our problem is also generalized to consider general communication cost that captures properties, such as energy consumption, bandwidth or storage space, and general latency penalty that models considerations, such as the ineffectiveness of delayed information, dissatisfaction of users, and incurred economic loss. Our problem can be applied to a broad range of applications.

In the literature \cite{Borodin:1998}, {\em competitive ratio} is a common metric to compare the performance of an online algorithm against an optimal offline algorithm. In our context, we compare the performance of a {\em distributed} online algorithm against a {\em centralized} optimal offline algorithm. We provide simple threshold-based competitive distributed online algorithms in various settings for our problem that match with the theoretical lower bounds of competitive ratios in order of magnitude. Moreover, we observe that our algorithms can produce satisfactory performance in both simulations and practical testbed.

%Distributed online algorithms are a burgeoning area, which is vital to many applications in networking and distributed systems \cite{Borodin:1998}. Our study concerns a general problem, paving the way for tackling a wide range of problems involving distributed online algorithms.

{\bf Outline:} The problem is formulated in Sec.~\ref{sec:model}. Simple online algorithms for various intercommunication settings are presented in Secs.~\ref{sec:whc}-\ref{sec:net}, which match with the lower bounds of competitive ratios in Sec.~\ref{sec:lb}. We performed simulations to corroborate the performance in Sec.~\ref{sec:sim}. We also implemented our algorithms in a practical sensor network testbed, with observations discussed in Sec.~\ref{sec:impl}. Some proofs are deferred to the technical report \cite{CKAToN:2016}, unless stated otherwise.

%\vspace{-5pt}
\section{Related Work} \label{sec:related}

%\vspace{-5pt}
\subsection{Online Algorithms}

Online algorithms are critical for a wide range of problems \cite{Borodin:1998, CZCTSG:2016, LTCCL:2013} involving uncertain input and timely decisions, by making decisions and generating outputs instantaneously over time, based on only the part of the input that have seen so far, without knowing the rest in the future.  
%The performance of online algorithms is usually evaluated using competitive analysis. The \textit{competitive ratio }\cite{Borodin:1998} of an online algorithm is defined as the worst-case ratio between the cost of the solution obtained by the online algorithm versus that of an optimal offline solution obtained by knowing the all input sequence in the future (e.g., with the presence of an omniscient oracle). 
Online algorithms have several benefits. First, they do not require a-prior or stochastic knowledge of the inputs, which is robust in any uncertain (even adversarial) environments, such as in dynamic ad hoc networks. Second, online algorithms are often simple decision-making mechanisms, without relying on future information prediction. Third, online algorithms give a fundamental characterization of the problems, which are useful to benchmark more sophisticated decision-making mechanisms. 

In this paper, we study a {\em distributed} online algorithmic problem, with multiple online decision-makers. %This class of problem is relatively new, and has only few prior results in the literature \cite{Aspnes1998competitive, Tziritas:2013}. 
This paper also extends the online decision problem in \cite{AftabChau:2013}, which considers a single sensor for temperature control in an online fashion. 

\subsection{Dynamic TCP Acknowledgement}

Dynamic TCP ACK problem \cite{Dooly:2001,Karlin:2001,Buchbinder:2007} is equivalent to a subclass of our problem (i.e., the one-system case). There are a stream of packets arriving at a destination. The packets must be acknowledged in order to notify the sender that the transmission was successful. 
However, it is possible to simultaneously acknowledge multiple packets using a single ACK packet. The delayed acknowledgement reduces the number of the ACKs, but might incur excessive latency to the TCP connection and interfere with the TCP's congestion control mechanism. 
The problem is to find a solution with the minimum cost of the total number of ACKs sent and the latency cost incurred by delayed ACKs.  Dooly et al. \cite{Dooly:2001} formulated this problem, and devised a 2-competitive deterministic algorithm. %\cite{Karlin:2001} achieved $\frac{e}{e-1}$-competitive ratio in expectation using a randomized online algorithm, which was also achieved using a primal-dual linear programming approach in \cite{Buchbinder:2007}. 
%Here, we will rely on the approach in \cite{Buchbinder:2007}.
Dynamic TCP ACK problem may be viewed as a recurrent version of ski-rental problem \cite{Lotker:2008}. Despite of many variants of ski-rental problem in the literature \cite{KhanaferKP13,Buchbinder:2007}, there lack results about distributed online decision-makers.

\subsection{Information Aggregation and Multicast Acknowledgement}

Information aggregation is a useful technique for in-situ sensing, which has been studied extensively in the literature \cite{FRWZ07survey}. The prior literature considered various system solutions, such as MAC and routing protocols that incorporate aggregation, aggregation functions that merge information from multiple sources, and representations of data to facilitate aggregation. In contrast, this paper considers the online decision-making mechanisms for information aggregation, which may be employed in tandem with the prior system solutions.

On the other hand, a closely related problem to our work is Multicast Acknowledgement Problem (MAP) \cite{BKV12agg,KNR02agg}. In MAP, a root sends data packets along a multicast tree, which has unreliable transmission links. Each node in the multicast tree are required to reply ACKs to the root for its received data packets. Rather than transmitting an ACK immediately, each node can delay its ACKs, by aggregating the ACKs for multiple data packets and the ACKs from the descendant nodes in the multicast tree. The key differences of MAP with our problem are that we do not require an ACK from every node (but only $K$ ACKs from all the receivers), and we consider non-hierarchical network such that every node can communicate with base station (unlike the tree hierarchy in MAP). The differences lead to different characterizations of competitive ratios of online algorithms.
We also note that there is a recent extension of MAP to consider bi-criteria optimization of communication cost and latency separately, rather than by a weighted objective function \cite{KMSV07agg,BKMSSV06agg}.

%\vspace{-5pt}
\section{Problem Formulation} \label{sec:model}

\subsection{Main Components}
This paper considers a general model of a group of distributed systems, which make online decisions for reporting certain correlatively observed events in a distributed manner. 

The model is consisted of the following components:
\begin{itemize}

\item {\bf Systems}:  A group of $N$ systems, indexed by set ${\cal N}$, are tasked with reporting certain events to a base station.

\item {\bf Events}: A sequence of events of interest, indexed by a finite set ${\cal M}$, appear over a continuous time horizon. For each event $j \in {\cal M}$, let its appearance time be ${\tt t}^j$. To model the observation by a system, let $w_i^j$ represent an abstract real-valued metric of observation for event $j$ at system $i$. If event $j$ is not observed by $i$, then $w_i^j = 0$. Denote the tuple of measurement of $j$ at $i$ by $\sigma_i^j \triangleq ({\tt t}^j, w_i^j)$, and the measurements at all systems by $\sigma \triangleq (\sigma_i^j)_{i \in {\cal N}, j \in{\cal M}}$. We assume that there is only one event occurred at each $t$, because of continuous time.

\item {\bf Reports}: For each event, at least $K (\le N)$  report messages from $K$ systems in ${\cal N}$ that observe the event (i.e., $w_i^j > 0$) are required by the base station. General content of a report message is allowed. For instance, a report message for notification may only include a Boolean flag if an event is observed. A report message may also contain more detailed information, such as the appearance times and measurement values of the observed events. This paper only considers the abstract communication cost associated with the report messages.

\end{itemize}

\smallskip

\begin{example}
There are a few examples of the model:
\begin{itemize}

\item {\bf Anycast ACKs}: An event is a reception of a packet $j$ at an anycast receiver $i$. This is an extension of the classical (unicast) TCP ACK problem to a distributed (anycast) setting. Let $K =1$, and $w_i^j =  1$ if packet $j$ is received by receiver $i$. A report message is an ACK packet.

\item {\bf In-situ Sensing}:  An event is an observation of a substance source $j$ at a sensor $i$, and $w_i^j$ is the observed concentration level of source $j$ at sensor $i$.  For example, in \cite{AftabChau:2013}, $w_i^j$ is the temperature fluctuation, when a heat source appears. A report message is a measurement report from a sensor. Sometimes, multiple report messages are required. For instance, tracing the location of the source may require the measurements from $K \ge 3$ sensors.

\item {\bf Distributed Caching}: An event is an object $j$ being cached at a distributed server $i$, and $w_i^j$ represents the size of object $j$, as cached by server $i$. A report message is a hash file of the cached objects from a server. The hash files will be distributed among other servers for notifying the presence of cached objects at other servers. For reliability, it is desirable to cache each object with at least $K$ servers, if available.

\end{itemize}
\end{example}

\subsection{Problem Definition} 

The systems in ${\cal N}$ make decisions whether to report the observed events at time $t$. Let the decisions of system $i$ be a sequence ${\cal R}_i = \big(r_{i,k},(x_k(i,j))_{j \in {\cal M}}\big)_{k = 1}^{\eta_i}$, where $r_{i,k}$ is the reporting time of the $k$-th report message, and $x_k(i,j)\in \{0, 1\}$ is an indicator variable whether measurement $\sigma_i^j$ is included in the $k$-th report message. Let $\eta_i$ be the total number of report messages from $i$. 
Let $\gamma^j(K)$ be the reporting time of the $K$-th report\footnote{
If event $j$ is never reported for $K$ times, then $\gamma^j(K) = \infty$. Without loss of generality, it is assumed that each event must trigger non-zero measurement values for at least $K$ systems.} from one of the observed systems for event $j$. Namely,
\begin{equation}
\gamma^j(K) \triangleq K^{\rm th}\mbox{-}{\rm smallest}\{ r_{i,k} \mid x_{k}(i,j) = 1 \wedge w_i^j > 0\}
\notag
\end{equation}
where $K^{\rm th}\mbox{-}{\rm smallest}$ selects the $K$-th smallest item from a set of numerical items, with a deterministic tie-breaking rule.

Denote the set of measurements included in the $k$-th report message of system $i$ by $\sigma_{i,k} \triangleq \{ \sigma_i^j:  x_k(i,j) = 1 \}$.

The goal of $K$-report {\bf D}istributed {\bf I}nformation {\bf A}ggregation problem ({\rm $K$-DIA}) is to minimize a global objective function of the communication cost of all report messages and the associated latency penalty for the $K$-th report messages, as defined as follows:
\begin{eqnarray} \label{eqn:DIA-def}
(K{\rm \mbox{-}DIA})  & & \min_{{\cal R}_i: i \in {\cal N}}  
\sum_{i \in {\cal N}} \Big(\rho \cdot \sum_{k=1}^{\eta_i}{\cal C}(\sigma_{i,k}) +   \\
 & & \qquad\qquad\quad \ (1-\rho) \cdot \sum_{j \in {\cal M}} {\cal L}(\sigma_i^j,\gamma^j(K))\Big) \quad  \notag
\end{eqnarray} 
\begin{eqnarray} \label{eqn:DIA-con}
\mbox{s.t.}  & & {\tt t}^j \le r_{i,k} \mbox{\ if\ } x_k(i,j) = 1,  \mbox{\ for all\ } j \in {\cal M}
\end{eqnarray} 
The notations in the objective are explained as follows:
\begin{enumerate}

\item ${\cal C}({\sigma}_{i,k})$ represents the abstract {\em communication cost} of a report message containing the set of measurements ${\sigma}_{i,k}$. The communication cost can reflect various properties, such as energy consumption, bandwidth or storage space of the report messages. We consider a general communication cost function. For example, if the communication cost depends on only the number of messages (e.g., for notification), then ${\cal C}(\sigma_{i,k}) = 1$. If the communication cost depends on the binary representation of the total measurement (e.g., for data archiving), then
\begin{equation}
{\cal C}(\sigma_{i,k}) = C_1 \cdot \log\Big(\big[\sum_{j\in{\cal M}: {\sigma}_i^j \in {\sigma}_{i,k}} w_i^j \big]_{\underline{w}}^{\bar{w}} \Big) + C_0
\end{equation}
where $[\cdot]_{\underline{w}}^{\bar{w}} \triangleq \min(\max(\cdot, \underline{w}), \bar{w})$ is an operator that rounds the measurement value within the range $[\underline{w}, \bar{w}]$.

Generally, ${\cal C}(\sigma_{i,k})$ is assumed to be non-decreasing in each parameter $w_i^j$, and {\em sub-additive}, such that
\begin{equation}
{\cal C}(\sigma_{i,k} \cup \sigma'_{i,k}) \le {\cal C}(\sigma_{i,k}) + {\cal C}(\sigma'_{i,k})
\end{equation}
Sub-additivity reflects the benefit of overhead reduction by aggregating multiple report messages into a single one. Suppose the communication cost per report is bounded by ${\cal C}_{\min} \le {\cal C}(\sigma_{i,k}) \le {\cal C}_{\max}$. Let $\alpha \triangleq \frac{{\cal C}_{\max}}{{\cal C}_{\min} }$, where $\alpha \ge 1$.

\item ${\cal L}(\sigma_i^j,\gamma^j(K))$ represents the {\em latency penalty} that captures the adverse effect when the $K$-th report message of event $j$ is delayed until $\gamma^j(K)$. The latency penalty can model factors, such as the ineffectiveness of delayed information, dissatisfaction of users, and incurred economic loss. We consider a general penalty function. For example, the penalty function can be a linear function: 
\begin{equation}
{\cal L}(\sigma_i^j,\gamma^j(K)) = w_i^j \cdot (\gamma^j(K) - {\tt t}^j)
\end{equation}
Generally, ${\cal L}(\sigma_i^j,\gamma^j(K))$ is assumed to be non-decreasing in each $w_i^j$ and $\gamma^j(K)$, and ${\cal L}(\sigma_i^j,{\tt t}^j) = 0$ (because there is no delay) and ${\cal L}(\sigma_i^j,\gamma^j(K)) = 0$ if $w_i^j=0$ (because the event is not observed).

\item $\rho$ is a parameter for controlling the trade-off between the communication cost and latency. A larger value of $\rho$ favors the reduction in communication cost, whereas a smaller value of $\rho$ favors the reduction in latency.

\end{enumerate}

Note that there is a trade-off in information aggregation between the communication cost and latency penalty. {\rm $K$-DIA} is a generalization of the classical dynamic TCP ACK problem \cite{Dooly:2001} by considering information aggregation from multiple sources.

For the convenience of illustration, we also define a sub-problem called {\rm $1$-sDIA}, which particularly captures the anycast ACK problem:
\begin{eqnarray} \label{eqn:sDIA-def}
(1{\rm \mbox{-}sDIA}) & & \min_{{\cal R}_i: i \in {\cal N}}  
\sum_{i \in {\cal N}} \Big(\rho \cdot \sum_{k=1}^{\eta_i}{\cal C}(\sigma_{i,k}) + \\
 & & \qquad\qquad\quad \ (1-\rho) \cdot \sum_{j \in {\cal M}} {\cal L}(\sigma_i^j,\gamma^j(1))\Big) \quad  \notag
\end{eqnarray} 
where ${\cal C}(\sigma_{i,k}) = 1$ and ${\cal L}(\sigma_i^j,\gamma^j(1)) = w_i^j \cdot (\gamma^j(1) - {\tt t}^j)$, and
\begin{equation}
\gamma^j(1) = \min_{i \in {\cal N}, k \in \{1, ..., \eta_i\}} \{r_{i,k} \mid  r_{i,k} \ge {\tt t}^j \wedge w_i^j > 0 \} \notag
\end{equation} 
Note that the classical TCP ACK problem is the one-system case (i.e., $|{\cal N}|=1$) of {\rm $1$-sDIA}.

%\vspace{-5pt}
\subsection{Distributed Online Algorithms}

Given all the inputs $\sigma = ({\tt t}^j, (w_i^j)_{i\in{\cal N}})_{j\in {\cal M}}$ in advance, {\rm $K$-DIA} can be solved in an offline manner. However, the measurements $\sigma$ are revealed gradually over time in practice. Furthermore, in multi-system setting, each system initially only knows its local measurements $(\sigma_i^j)_{j \in {\cal M}}$,  without prior knowledge of the local observations at others. Thus, there is incomplete information at each system about the inputs. The incomplete information presents a unique challenge in the multi-system setting, as compared to the one-system setting (i.e., dynamic TCP ACK problem).

We will provide {\em distributed online} algorithms for solving {\rm $K$-DIA}. More specifically, we focus on {\em deterministic} distributed algorithms with the following characteristics:
\begin{enumerate}

\item The decisions ${\cal R}_i$ at each system follow a set of pre-defined deterministic rules. 
%\vspace{-5pt}

\item The measurements $(\sigma_i^j)_{j \in {\cal M}}$ are observed locally by each system $i$ only, and cannot be revealed to other systems without involving communication. 
%\vspace{-5pt}

\item If there involves any communication among the systems, then this also incurs a cost per transmission, which may be as {\em costly} as communicating a report message to the base station. Thus, it may be more efficient by communicating a report message to the base station, and letting other systems overhear it, if overhearing is possible.

\end{enumerate}
An algorithm is called {\em online}, if the decision of at the time $t$ only depends on the information available before or at $t$. Namely, the decision of producing a report message at time $t$ is only based on the measurements $(\sigma_i^j)_{j: {\tt t}^j \le t, i \in {\cal N}}$. 

A simple class of online algorithms are threshold-based algorithms.
First, define the {\em accumulative latency penalty} at system $i$ for an interval $(t_1, t_2]$ by
\begin{equation} \label{eqn:lat-def}
{\tt lat}_i(t_1, t_2) \triangleq \sum_{j \in {\cal M} : {\tt t}^j \ge t_1} {\cal L}(\sigma_i^j, t_2)
\end{equation} 
and the {\em accumulative communication cost} by 
\begin{equation} \label{eqn:com-def}
{\tt com}_i(t_1, t_2) \triangleq {\cal C}\big(\{ \sigma_i^j:  t_1 < {\tt t}^j \le t_2 \}\big)
\end{equation} 
In a {\em threshold-based} algorithm, each report message includes all the unreported events that are observed locally. A report message is produced, if the ratio between the accumulative latency penalty and the accumulative communication cost since the last report message exceeds a pre-defined threshold $\theta$. For example, in {\rm $1$-DIA}, the reporting time of the $k$-th report message at system $i$ is obtained by
\begin{equation}
r_{i,k} = \min\Big\{t : \frac{{\tt lat}_i(r_{i, k-1}, t)}{{\tt com}_i(r_{i, k-1}, t)}  \ge \theta \Big\}
\end{equation}
Threshold-based algorithm is online, since the accumulative latency penalty and accumulative communication cost depend on no future information. This paper studies the effectiveness of threshold-based algorithms.

%\vspace{-5pt}
\subsection{Intercommunication} \label{sec:intercom}

In addition to the report messages to the base station, extra intercommunication may be enabled among systems. We consider implicit intercommunication, such that one system may overhear the messages that are not intended for it (e.g., other report messages to the base station). The extent of intercommunication is characterized by the following cases:
\begin{enumerate}

\item {\bf No Intercommunication:}
There is no intercommunication among the systems. Namely, no system needs to communicate among themselves, nor can they overhear from each other. But the systems can communicate to the base station for reporting only. For example, in anycast over a wireline network, overhearing is difficult. 
%\vspace{-5pt}
%
\item {\bf Full Intercommunication:}
In contrast to no intercommunication, there is full intercommunication among the systems. For instance, a system can overhear report messages produced by any other systems and hence, possibly withholds its own report messages. This can be achieved, for instance, in a broadcast channel or by replicating the messages by the base station to other systems. For example, in wireless sensor networks, overhearing is possible.  An ideal setting of full intercommunication is {\em collision-free}, if no pair of systems transmits messages simultaneously\footnote{  {\em Collision-free} intercommunication can be implemented by several ways. For example: (1) Priority based: Given a pre-defined priority order, a system in lower priority is required to give way to those systems in higher priority by spending more time to sense the channel before transmission. (2) Random access: Before transmission, each system should wait for an exponentially distributed random interval before transmission. The probability of collision is negligible in continuous time.}. This can effectively eliminate redundant report messages produced at the same time. 
%\vspace{-5pt}
%
\item {\bf Partial Intercommunication:}
In practice, the extent of intercommunication is not prefect, which lies between the two preceding cases. We model the availability of intercommunication among systems by a {\em communication graph}, such that only the neighboring systems in the communication graph can engage in full intercommunication (including overhearing) with each other. %A communication graph reflects the physical limitations (e.g., obstacles) or logical constraints (e.g., packet filtering). 
\end{enumerate}

In the subsequent sections, {\rm $K$-DIA} will be studied under these cases of intercommunication.

\medskip

\begin{figure}[htb!] 
%\vspace{-5pt} 
\centering 
\includegraphics[scale=0.5]{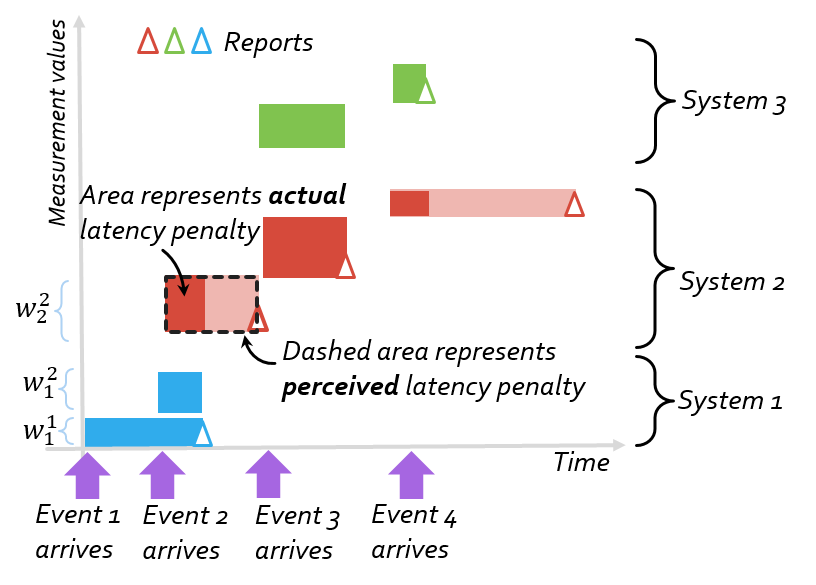}  
%\vspace{-5pt}
\caption{\small  An illustration of threshold-based algorithm for {\rm $1$-sDIA}. Four events arrive over time, whose measurement values to each system are plotted in $y$-axis. The rectangles represent the latency penalty, whereas the triangles represent the report messages.} \label{fig:chart0} 
%\vspace{-5pt}  
\end{figure} 

\begin{example}
This example considers {\rm $1$-sDIA}. The basic idea of the problem and the operation of threshold-based algorithm are illustrated in Fig.~\ref{fig:chart0}. In this example, system 2 cannot overhear others, but other systems can. Note that the communication cost in {\rm $1$-sDIA} is always one per message. The threshold-based algorithm is applied as follows:
\begin{enumerate}

\item All systems defer the report messages until the perceived accumulative latency penalty since the last report message exceeds a threshold. Each report message includes all unreported events that are observed locally.

\item If a system overhears a report message produced by another system, it needs not transmit a report message for those reported events. It resets the accumulative latency penalty, if all observed events have been reported.

\end{enumerate}
In Fig.~\ref{fig:chart0}, the actual accumulative latency penalty of each event is represented by the rectangles in deep colors. Since system 2 is not aware of the report messages produced by others, its perceived accumulative latency penalty may be more than the actual one, which is represented by the total area of deep and light colors (e.g., the rectangle in dashed line). In this example, five report messages are generated.
\end{example}

%\vspace{-5pt}
\subsection{Competitive Analysis and Competitive Ratio} 

In the literature of online algorithms, competitive analysis is a common framework to characterize the performance \cite{Borodin:1998}.

Denote the cost of online algorithm ${\cal A}$ by ${\tt Cost}({\cal A}[\sigma])$ and the cost of an optimal offline solution on input $\sigma$ by ${\tt Opt}(\sigma)$.
The {\em competitive ratio} for ${\cal A}$ is defined as the worst-case ratio between the cost of the online algorithm and that of an optimal offline solution, namely,
\begin{equation}
{\tt CR}({\cal A}) \triangleq \max_{\sigma} \frac{{\tt Cost}({\cal A}[\sigma])}{{\tt Opt}(\sigma)}
\end{equation}
${\cal A}$ is called $c$-competitive, if ${\tt CR}({\cal A}) = c$. 

In the context of distributed online algorithms, we compare with the (centralized) optimal offline solution, which knows all the inputs in advance without the need for intercommunication. 

In this paper, we show that threshold-based algorithms with proper thresholds can produce competitive performance for solving {\rm $K$-DIA} that are close to the theoretical lower bounds of any deterministic online algorithms. The main results are summarized in Table~\ref{tab:res}.

\begin{table}[htb!] \centering
{ 
\begin{tabular}{@{\ }c@{\ }|@{\ }c@{\ }|@{\ }c@{\ }}
  \hline  \hline 
   Competitive Ratio & \ No Intercommunication & Full Intercommunication \\ 
  \hline                       
  \multirow{2}{*}{Threshold-based} & \multirow{2}{*}{$\frac{\alpha N}{K}+1$}  & \multirow{2}{*}{$\frac{\sqrt{(K-\alpha)^2 + 4 \alpha K N} + \alpha + K}{2 K}$}\\
  &  & \\
  \hline 
  Lower bound  & \multirow{2}{*}{$\alpha(N-1)+\frac{3}{2}$} & \multirow{2}{*}{$\frac{\sqrt{N}}{4}$}\\
  ($K=1$) &  & \\
  \hline  \hline 
  Competitive Ratio & \multicolumn{2}{@{}c@{}}{Partial Intercommunication} \\
    \hline 
  \multirow{2}{*}{Threshold-based} & \multicolumn{2}{@{}c@{}}{$\frac{\sqrt{(\alpha {\rm x} - K)^2 + 4 \alpha K N} + \alpha {\rm x} + K}{2 K} $} \\
   & \multicolumn{2}{@{}c@{}}{where $1 \le {\tt x} \le N$ is a network parameter} \\
    \hline  \hline 
\end{tabular} } 
%\vspace{-10pt}
\caption{\small   Competitive ratios of threshold-based algorithms, and the theoretical lower bounds of any deterministic online algorithms.} \label{tab:res} 
%\vspace{-5pt}
\end{table}

\section{No Intercommunication} \label{sec:woc}

This section considers the simplest setting with no intercommunication among the systems. 
Threshold-based algorithm ${\cal A}_{\rm thb}$ (Algorithm~\ref{alg:woc}) produces a report message independently at each system $i$ when the ratio between the perceived accumulative latency penalty and the accumulative communication cost since the last report message exceeds threshold $\theta$. Note that for each event, ${\cal A}_{\rm thb}$ produces report messages at all the systems that observe non-zero measurement values. Hence, every event triggers $N (\ge K)$ report messages, and this produces a feasible solution for {\rm $K$-DIA}.

\begin{algorithm}
\caption{\small  ${\cal A}_{\rm thb}[\theta, t_{\rm now}, (\sigma_i^j)_{j:{\tt t}^j\le t_{\rm now}}]$} \label{alg:woc}
{\small  \begin{algorithmic}[1]
\Require Threshold $\theta$; current time $t_{\rm now}$; known measurements $(\sigma_i^j)_{j:{\tt t}^j\le t_{\rm now}}$
\Statex \hspace{-15pt} {\bf Global Init.:} $x_k(i,j) \leftarrow 0$ for all $j \in {\cal M}, k \leftarrow 0, r_{i,k} \leftarrow 0$
\vspace{3pt} \hrule \vspace{3pt}
%\Statex \Comment{Find the last notifying time}
%\State $r_{\rm pre} \leftarrow \max\{ t \in {\cal R}_i\}$ \Comment{{\em Find the last reporting time}}
\If {$\frac{{\tt lat}_i(r_{i,k}, t_{\rm now})}{{\tt com}_i(r_{i,k}, t_{\rm now})} \ge \theta$} \Comment{{\em Condition on threshold} $\theta$}
%\Statex \Comment{Generate a notification at time $t_{\rm now}$}
\State $k \leftarrow k + 1$\Comment{{\em Next report message}}
\State $x_k(i,j) \leftarrow 1$ for all observed event $j$ 
\State $r_{i,k} \leftarrow t_{\rm now}$ 
\EndIf 
\end{algorithmic}} 
\end{algorithm}

Denote by ${\cal A}_{\rm thb}[\theta, \sigma]$ the threshold-based algorithm with threshold $\theta$ on input $\sigma$. By a slight abuse of notation, we also denote by ${\tt Opt}(\sigma)$ the optimal offline solution on input $\sigma$.  

First, we consider {\rm $1$-DIA} of at least one report message per event. Given input $\sigma$, let ${x}_k^\ast(j) \in \{0, 1\}$ be the indicator variable whether event $j$ is included in the $k$-th report message by ${\tt Opt}(\sigma)$. The set of events ${\cal M}$ can be partitioned into a set of segments ${\cal S}^\ast$, grouped by the corresponding report messages, such that ${\tt S}_k \in {\cal S}^\ast$ is defined by
\begin{equation} 
 {\tt S}_k	 \triangleq \{j \in {\cal M} \mid {x}_k^\ast(j) = 1\}
\end{equation}
For each segment ${\tt S} \in {\cal S}^\ast$, denote the sub-sequence of measurements in ${\tt S}$ by $\sigma_{\tt S} \triangleq (\sigma_i^j)_{i \in {\cal N}, j \in {\tt S}}$. 
The competitive ratio can be obtained considering each $\sigma_{\tt S}$.

\smallskip

\begin{figure}[htb!] 
\hspace{-10pt}
%\centering 
\includegraphics[scale=0.47]{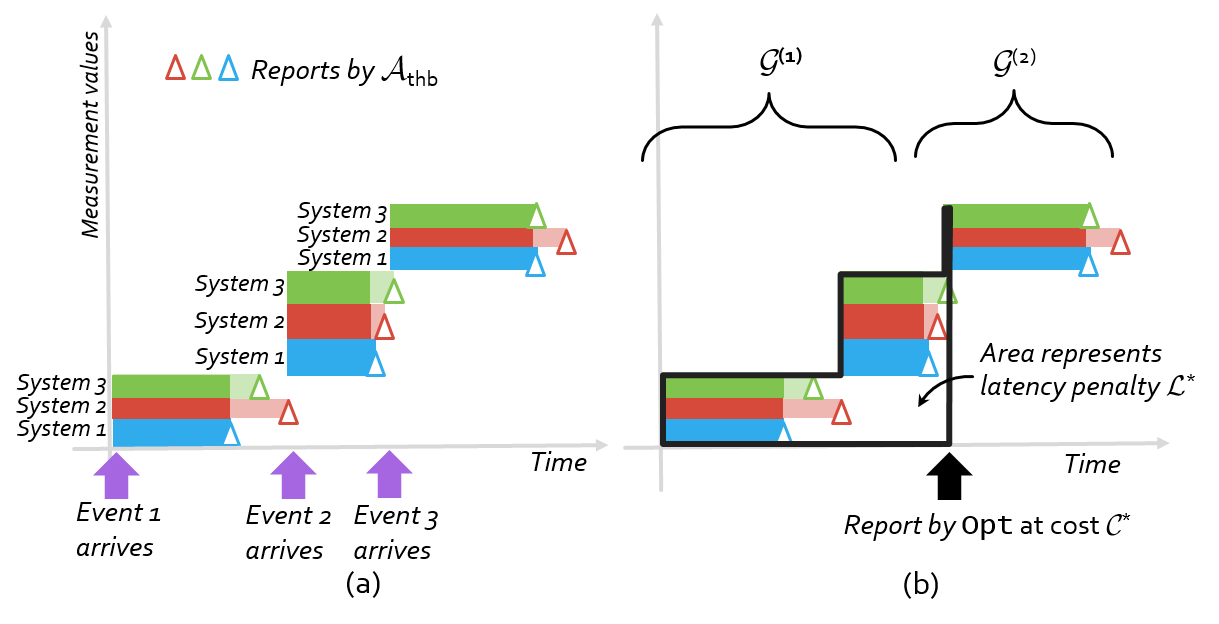}  
%\vspace{-25pt}
\caption{\small   Consider {\rm $1$-sDIA}. (a) ${\cal A}_{\rm thb}$ produces a report message independently at each system. (b) ${\tt Opt}$ produces a report message in the end of each segment ${\tt S} \in {\cal S}^\ast$ with communication cost ${\cal C}^\ast$ and latency penalty ${\cal L}^\ast$ (i.e., the polygon area in black border). The set of measurements is divided into two groups ${\cal G}^{(1)}$, ${\cal G}^{(2)}$.\label{fig:chart1}}  
%\vspace{-10pt}
\end{figure} 

\begin{customthm}{1} \label{thm:woc.alg}
For {\rm $1$-DIA}, let threshold $\theta = {\textstyle\frac{\rho}{\alpha N(1-\rho)}}$ for every system $i \in {\cal N}$, the competitive ratio of ${\cal A}_{\rm thb}$ is ${\tt CR}({\cal A}_{\rm thb}[{\textstyle\frac{\rho}{\alpha N(1-\rho)}}]) = \alpha N+1$.
\end{customthm}

\begin{IEEEproof}
The optimal offline solution ${\tt Opt}(\sigma)$ determines a set of segments ${\cal S}^\ast$.
Note that ${\tt Opt}(\sigma) = \sum_{{\tt S} \in {\cal S}^\ast} {\tt Opt}(\sigma_{\tt S})$. By Lemma~\ref{lem:subadd} (see \cite{CKAToN:2016}), we obtain
\begin{equation}
{\tt Cost}({\cal A}_{\rm thb}[\theta, \sigma]) \le \sum_{{\tt S} \in {\cal S}^\ast} {\tt Cost}({\cal A}_{\rm thb}[\theta, \sigma_{\tt S}])
\end{equation}
Since $\frac{\sum_{l} a_l}{\sum_{l} b_l} \le \max_l\{\frac{a_l}{b_l}\}$ for any positive $a_l, b_l$, we obtain
\begin{equation} \hspace{-5pt}
\frac{ {\tt Cost}({\cal A}_{\rm thb}[{\textstyle\frac{\rho}{\alpha N(1-\rho)}}, \sigma]) }{{\tt Opt}(\sigma)} \le \max_{{\tt S} \in {\cal S}^\ast}
\frac{ {\tt Cost}({\cal A}_{\rm thb}[{\textstyle\frac{\rho}{\alpha N(1-\rho)}}, \sigma_{\tt S}]) }{{\tt Opt}(\sigma_{\tt S})}
\end{equation}
The rest of proof aims to find an upper bound for $\frac{ {\tt Cost}({\cal A}_{\rm thb}[{\textstyle\frac{\rho}{\alpha N(1-\rho)}}, \sigma_{\tt S}]) }{{\tt Opt}(\sigma_{\tt S})}$ for any given ${\tt S}$.

Recall that ${\tt Opt}(\sigma_{\tt S})$ produces a report message in the end of ${\tt S}$. Let the optimal offline cost be ${\tt Opt}(\sigma_{\tt S}) = \rho {\cal C}^{\ast} + (1-\rho) {\cal L}^{\ast}$, where ${\cal C}^{\ast}$ and ${\cal L}^{\ast}$ are the communication cost and latency penalty of ${\tt Opt}(\sigma_{\tt S})$ respectively.

Next, we divide the set of measurements $\{\sigma_i^j\}_{ j \in {\tt S}, i \in {\cal N} }$ into two groups ${\cal G}^{(1)}, {\cal G}^{(2)}$:
\begin{itemize}

\item ${\cal G}^{(1)}$ is the set of measurements in $\{\sigma_i^j\}_{ j \in {\tt S}, i \in {\cal N} }$, such that ${\cal A}_{\rm thb}[{\textstyle\frac{\rho}{\alpha N(1-\rho)}}]$ produces report messages before or at the same time as ${\tt Opt}(\sigma_{\tt S})$ does.

\item ${\cal G}^{(2)}$ is the set of measurements in $\{\sigma_i^j\}_{ j \in {\tt S}, i \in {\cal N} }$, such that ${\cal A}_{\rm thb}[{\textstyle\frac{\rho}{\alpha N(1-\rho)}}]$ produces report messages after ${\tt Opt}(\sigma_{\tt S})$ does. Note that one system can at most produce one report message for ${\cal G}^{(2)}$.
\end{itemize}
See Fig.~\ref{fig:chart1} for an illustration considering {\rm $1$-sDIA}.

Regarding to ${\cal A}_{\rm thb}[{\textstyle\frac{\rho}{\alpha N(1-\rho)}}]$, let the actual total latency penalty for ${\cal G}^{(1)}$ be ${\cal L}^{(1)}$, and the perceived total latency penalty be $\hat{\cal L}^{(1)}$. Note that ${\cal L}^{(1)} \le \hat{\cal L}^{(1)}$ because the systems cannot overhear the report messages from other systems. Let the total communication cost for ${\cal G}^{(1)}$ be ${\cal C}^{(1)}$. Similarly, we define ${\cal L}^{(2)}$, $\hat{\cal L}^{(2)}$ and ${\cal C}^{(2)}$ for ${\cal G}^{(2)}$.
Since ${\cal A}_{\rm thb}[{\textstyle\frac{\rho}{\alpha N(1-\rho)}}]$ is a threshold-based algorithm, it follows that 
\begin{equation}
\hat{\cal L}^{(1)} = {\frac{\rho}{\alpha N(1-\rho)}} {\cal C}^{(1)}, \qquad  \hat{\cal L}^{(2)} = {\frac{\rho}{\alpha N(1-\rho)}} {\cal C}^{(2)}
\end{equation}

First, consider ${\cal G}^{(1)}$. Note that $\hat{\cal L}^{(1)} \le {\cal L}^\ast$, because ${\cal G}^{(1)}$ is the set of measurements that ${\cal A}_{\rm thb}[{\textstyle\frac{\rho}{\alpha N(1-\rho)}}]$ produces report messages before or at the same time as ${\tt Opt}(\sigma_{\tt S})$ does, and the latency penalty function is non-decreasing in the reporting time. See Fig.~\ref{fig:chart1} (b) for an illustration.
Hence,
\begin{equation}
{\cal C}^{(1)}  = {\frac{\alpha N(1-\rho)}{\rho}} \hat{\cal L}^{(1)} \le {\frac{\alpha N(1-\rho)}{\rho}} {\cal L}^\ast
\end{equation}
Therefore, the total cost of ${\cal A}_{\rm thb}$ for ${\cal G}^{(1)}$ is bounded by
\begin{eqnarray}
& & {\tt Cost}({\cal A}_{\rm thb}[{\textstyle\frac{\rho}{\alpha N(1-\rho)}}, {\cal G}^{(1)}]) \\
& = &   \rho {\cal C}^{(1)} + (1-\rho)  {\cal L}^{(1)}  \le  (1-\rho)(\alpha N + 1) {\cal L}^\ast
\end{eqnarray}

Second, consider ${\cal G}^{(2)}$.
Let the number of report messages by ${\cal A}_{\rm thb}[{\textstyle\frac{\rho}{\alpha N(1-\rho)}}]$ for ${\cal G}^{(2)}$ be $\eta^{(2)}$. Note that $\eta^{(2)} \le N$, because there are at most $N$ systems. Since the communication cost per message is bounded by $\alpha$, we obtain $\alpha N {\cal C}^{\ast}  \ge  {\cal C}^{(2)}$.
Thus,
\begin{equation}
{\cal L}^{(2)} \le  \hat{\cal L}^{(2)} = {\frac{\rho}{\alpha N(1-\rho)}} {\cal C}^{(2)} \le {\frac{\rho}{1-\rho}} {\cal C}^\ast 
\end{equation} 
Therefore, the total cost of ${\cal A}_{\rm thb}$ for ${\cal G}^{(2)}$ is bounded by
\begin{eqnarray}
& & {\tt Cost}({\cal A}_{\rm thb}[{\textstyle\frac{\rho}{\alpha N(1-\rho)}}, {\cal G}^{(2)}]) \\
& = &   \rho {\cal C}^{(2)} + (1-\rho)  {\cal L}^{(2)}  \le  \rho(\alpha N + 1) {\cal C}^\ast 
\end{eqnarray}
To sum up, the competitive ratio is obtained by
\begin{eqnarray}
& & \frac{ {\tt Cost}({\cal A}_{\rm thb}[{\textstyle\frac{\rho}{\alpha N(1-\rho)}}, \sigma_{\tt S}]) }{{\tt Opt}(\sigma_{\tt S})}  \\
& \le & \frac{(1-\rho) (\alpha N + 1) {\cal L}^{\ast} + \rho(\alpha N + 1){\cal C}^{\ast}}{\rho {\cal C}^{\ast} + (1-\rho) {\cal L}^{\ast}}  =  \alpha N + 1
\end{eqnarray}
\end{IEEEproof}

{\bf Remark:} Theorem~\ref{thm:woc.alg} naturally generalizes the 2-competitive deterministic online algorithm for the classical TCP ACK problem in \cite{Dooly:2001} (i.e., the one-system case of {\rm $1$-sDIA}). 
We will show in Sec.~\ref{sec:lb} that the competitive ratio is indeed optimal in terms of order of magnitude among any deterministic distributed online algorithms, as the lower bound is $\Omega(N)$.

\smallskip

\begin{customthm}{2} \label{thm:woc.alg.k}
For {\rm $K$-DIA}, let threshold $\theta = {\textstyle\frac{K \rho}{\alpha N(1-\rho)}}$ for every system $i \in {\cal N}$, the competitive ratio of ${\cal A}_{\rm thb}$ is ${\tt CR}({\cal A}_{\rm thb}[{\textstyle\frac{K \rho}{\alpha N(1-\rho)}}]) = \frac{\alpha N}{K}+1$.
\end{customthm}

{\bf Remark:}  %The larger value of $K$, the higher communication cost it incurs. Thus, the optimal offline solution becomes biased toward the reduction in communication cost. 
The competitive ratio of ${\cal A}_{\rm thb}$ diminishes with $K$, because ${\tt Opt}$ has increased communication cost, while ${\cal A}_{\rm thb}$ produces the same communication cost independent of $K$.

\section{Full Intercommunication} \label{sec:whc}

This section considers full intercommunication between any pair of systems. 
We study the extent of intercommunication can impact on the competitive ratio of online decision making. With the help of collision-free intercommunication, the competitive ratio can be improved from $O(N)$ to $O(\sqrt{N})$, which matches the lower bound $\Omega(\sqrt{N})$ in Sec.~\ref{sec:lb}. 

Threshold-based algorithm ${\cal A}^{{\rm itc.}K}_{\rm thb}$ (Algorithm~\ref{alg:whc}) relies on overhearing between every pair of systems. In an ideal setting, we suppose that there is a priority based collision-free intercommunication mechanism. The local algorithms at systems are invoked sequentially following the priority order, without being invoked simultaneously. Each system $i$ produces a report message when the ratio between the perceived accumulative latency penalty and the accumulative communication cost since the last report message exceeds threshold $\theta$, similar to ${\cal A}_{\rm thb}$. However, each system can overhear the report messages from other systems. It will not produce a report message for those events that have been reported $K$ times. It also adjusts the accumulative latency penalty for the unreported events. 
${\cal A}^{{\rm itc.}K}_{\rm thb}$ ensures that each event is reported at least $K$ times, and hence produces a feasible solution for {\rm $K$-DIA}.

Let ${z}_i^j(t)$ be the indicator variable whether system $i$ has reported event $j$ at time $t$.
It is assumed that each event is identifiable in the report message (e.g., by the appearance time, as there is only one event at each time $t$). Because of the ability of overhearing report messages, the information $({z}_{i'}^j(t))_{i' \in{\cal N}}$ is shared among all the systems.

\smallskip

\begin{algorithm}
\caption{\small   ${\cal A}^{{\rm itc.}K}_{\rm thb}[\theta, t_{\rm now}, (\sigma_i^j)_{j:{\tt t}^j\le t_{\rm now}}, ({z}_{i'}^j(t_{\rm now}))_{i' \in{\cal N}}]$} \label{alg:whc}
{\small  \begin{algorithmic}[1] 
\Require Threshold $\theta$; current time $t_{\rm now}$; known measurements $(\sigma_i^j)_{j:{\tt t}^j\le t_{\rm now}}$; indicators of reported events ${z}_{i'}^j(t_{\rm now})$
\Statex \hspace{-15pt} {\bf Global Init.:} $x_k(i,j) \leftarrow 0$ for all $j \in {\cal M}, k \leftarrow 0, r_{i,k} \leftarrow 0$
\vspace{3pt} \hrule \vspace{3pt}
\For{each $j$ such that $r_{i,k} < {\tt t}^j \le t_{\rm now}$}
\State ${\rm cnt}_j \leftarrow \sum_{i' \in{\cal N}} {z}_{i'}^j(t_{\rm now})$ \Comment{{\em Count num. of report messages}}
\EndFor
\Statex \Comment{{\em Compute latency penalty for events reported $<$ K times}}
\State ${\tt lat} \leftarrow \displaystyle \sum_{j \in {\cal M} : r_{i,k} < {\tt t}^j \le t_{\rm now} \wedge {\rm cnt}_j < K} {\cal L}(\sigma_i^j, t_{\rm now})$
\Statex \Comment{{\em Compute communication cost for events reported $<$ K times}}
\State ${\tt com} \leftarrow {\cal C}\Big(\{ \sigma_i^j:  j \in {\cal M}, r_{i,k} < {\tt t}^j \le t_{\rm now} \wedge {\rm cnt}_j < K \}\Big)$
\Statex 
\If {$\frac{{\tt lat}}{{\tt com}} \ge \theta$} \Comment{{\em Condition on threshold}}
\State $k \leftarrow k + 1$\Comment{{\em Next report message}}
\State $x_k(i,j) \leftarrow 1$ for all observed event $j$ 
\State $z^j_i(t_{\rm now}+\epsilon) \leftarrow 1$ for all reported event $j$, and any $\epsilon$
\State $r_{i,k} \leftarrow t_{\rm now}$ 
\EndIf 
\end{algorithmic}}
\end{algorithm}

\begin{figure}[htb!] 
%\vspace{-10pt}
\hspace{-10pt} 
\includegraphics[scale=0.47]{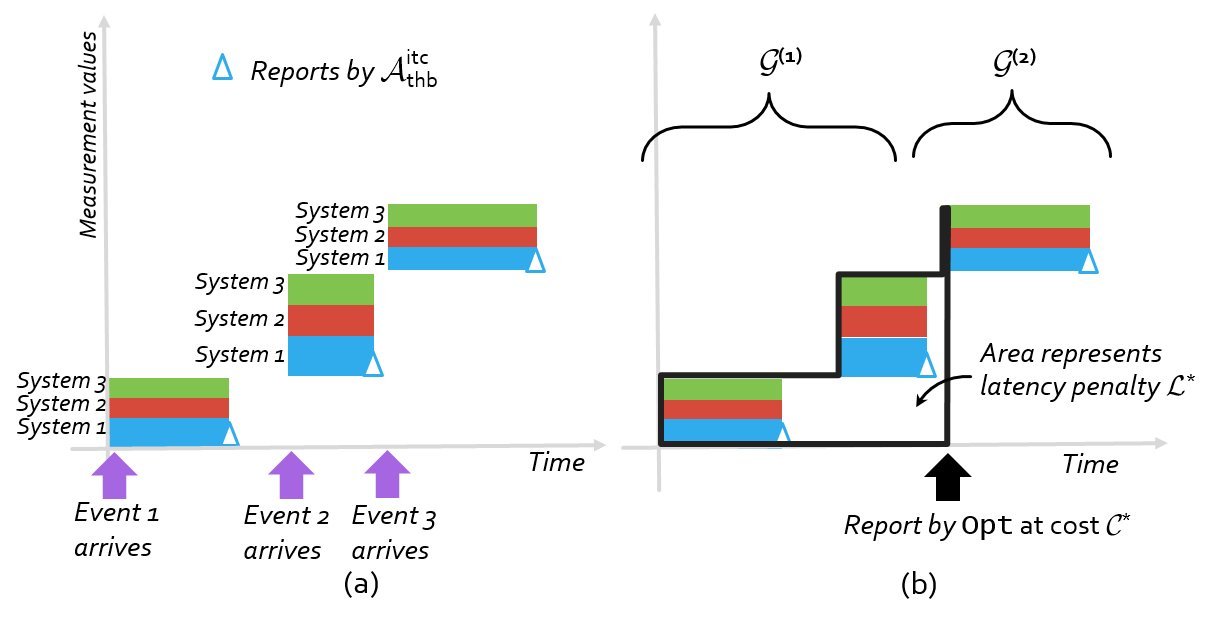}  
%\vspace{-25pt}
\caption{\small   (a) Considering {\rm $1$-DIA}, ${\cal A}^{{\rm itc.}1}_{\rm thb}$ produces one report message per event. (b) ${\tt Opt}$ produces a report message in the end of each segment ${\tt S} \in {\cal S}^\ast$ with communication cost ${\cal C}^\ast$ and latency penalty ${\cal L}^\ast$. The set of measurements is divided into two groups ${\cal G}^{(1)}$, ${\cal G}^{(2)}$.\label{fig:chart2}}  
%\vspace{-10pt}
\end{figure}

\begin{customthm}{3} \label{thm:whc.alg}
For {\rm $1$-DIA}, let threshold $\theta = {\textstyle\frac{\rho}{(1-\rho)\phi}}$ for every system $i \in {\cal N}$ where $\phi = \frac{\sqrt{(\alpha-1)^2 + 4 \alpha N}+\alpha-1}{2}$, the competitive ratio ${\tt CR}({\cal A}^{{\rm itc.}1}_{\rm thb}[{\textstyle\frac{\rho}{(1-\rho)\phi}}]) = \frac{\sqrt{(\alpha-1)^2 + 4 \alpha N}+\alpha+1}{2}$.
\end{customthm}

\begin{IEEEproof}
As in Theorem~\ref{thm:woc.alg}, we define ${\cal S}^\ast$, ${\cal G}^{(1)}$, ${\cal G}^{(2)}$, ${\cal C}^{\ast}$, ${\cal L}^{\ast}$, ${\cal C}^{(1)}$, ${\cal L}^{(1)}$, ${\cal C}^{(2)}$, and ${\cal L}^{(2)}$. The competitive ratio is obtained with respect to each $\sigma_{\tt S}$ for ${\tt S} \in {\cal S}^\ast$. 

First, consider ${\cal G}^{(1)}$. Note that because of overhearing, the perceived latency penalty is identical with the actual latency penalty. Hence, ${\cal L}^{(1)} \le {\cal L}^\ast$. See Fig.~\ref{fig:chart2} for an illustration.

Note that ${\cal A}^{{\rm itc.}1}_{\rm thb}$ produces at most one report message per event. Given ${\cal L}^\ast$, the maximum value of ${\cal C}^{(1)}$ is attained when each event is observed by only one system, while other $N-1$ systems incur zero latency. In this case, we obtain
\begin{equation}
{\cal C}^{(1)}  = {\frac{(1-\rho)\phi}{\rho}} {\cal L}^{(1)} \le {\frac{(1-\rho)\phi}{\rho}}  {\cal L}^\ast
\end{equation}
Therefore, the total cost of ${\cal A}^{{\rm itc.}1}_{\rm thb}$ for ${\cal G}^{(1)}$ is bounded by
\begin{equation}
{\tt Cost}({\cal A}^{{\rm itc.}1}_{\rm thb}[{\textstyle{\textstyle\frac{\rho}{(1-\rho)\phi}}}, {\cal G}^{(1)}]) 
 \le (1-\rho)(\phi + 1) {\cal L}^\ast
\end{equation}

Second, consider ${\cal G}^{(2)}$. We bound ${\cal L}^{(2)}$ by ${\cal C}^{(2)}$. Given ${\cal C}^{(2)}$, the maximum value of ${\cal L}^{(2)}$ is attained when all $N$ systems observe the same measurement value, while only one report message is generated by one of the systems (because of collision-free intercommunication). Hence,
\begin{equation}
 {\cal L}^{(2)} \le {\frac{N \rho}{(1-\rho)\phi}} {\cal C}^{(2)} 
\end{equation}
Since there are no simultaneous report messages with collision-free intercommunication, there is at most one report message for ${\cal G}^{(2)}$. Hence, $\alpha {\cal C}^{\ast} \ge {\cal C}^{(2)}$. Therefore, the total cost of ${\cal A}^{{\rm itc.}1}_{\rm thb}$ for ${\cal G}^{(2)}$ is bounded by
\begin{eqnarray}
& & {\tt Cost}({\cal A}^{{\rm itc.}1}_{\rm thb}[{\textstyle\frac{\rho}{(1-\rho)\phi}}, {\cal G}^{(2)}]) \\
& = &  \rho {\cal C}^{(2)} + (1-\rho) {\cal L}^{(2)} \le  \alpha \rho{\cal C}^{\ast}  + \alpha \rho \frac{N}{\phi} {\cal C}^{\ast}
\end{eqnarray}

To sum up, the competitive ratio is obtained by
\begin{eqnarray}
\displaystyle
& & \frac{ {\tt Cost}({\cal A}^{{\rm itc.}1}_{\rm thb}[{\textstyle\frac{\rho}{(1-\rho)\phi}}, \sigma_{\tt S}]) }{{\tt Opt}(\sigma_{\tt S})} \\
 & \le &
\displaystyle \frac{(1-\rho) (\phi + 1) {\cal L}^{\ast} + \alpha \rho(\frac{N}{\phi}+1){\cal C}^{\ast}}{\rho {\cal C}^{\ast} + (1-\rho) {\cal L}^{\ast}} \\
& = & \phi + 1 = \frac{\sqrt{(\alpha-1)^2 + 4 \alpha N}+\alpha+1}{2}
\end{eqnarray}
The last equality is because that $\phi = \frac{\sqrt{(\alpha-1)^2 + 4 \alpha N}+\alpha-1}{2}$ is the positive root of $\phi$ in $\phi + 1 = \alpha(\frac{N}{\phi}+1)$.
\end{IEEEproof}

\smallskip

{\bf Remark:} We note that it is possible to improve the competitive ratio of ${\cal A}^{{\rm itc.}1}_{\rm thb}$ by using heterogeneous thresholds. In the proof of Theorem~\ref{thm:whc.alg}, we assume the worst case where the system of maximum communication cost produces a report. One possible approach of improvement is to let the system of minimum communication cost to have a slightly smaller threshold $\theta_i$ than other systems, and hence, this system will be triggered to produce a report before others. Let 
$\theta_i = \theta + \epsilon\cdot{\cal C}(\sigma_{i,k})$,
where ${\cal C}(\sigma_{i,k})$ is communication cost of the next report, and $\epsilon$ is an arbitrarily small constant. Hence, the system with a higher communication cost will wait slightly longer. Thus, a report is always produced by the system of minimum communication cost. This is equivalent to having $\alpha = 1$ in Theorem~\ref{thm:whc.alg}, and we set
$\theta = {\textstyle\frac{\rho}{(1-\rho)\phi}}$, where $\phi = \sqrt{N}$. The competitive ratio of ${\cal A}^{{\rm itc.}1}_{\rm thb}$ becomes $\sqrt{N} + 1$.

\smallskip

\begin{customthm}{4} \label{thm:whc.alg.k}
For {\rm $K$-DIA}, let threshold $\theta = {\textstyle\frac{\rho}{(1-\rho)\phi}}$ for every system $i \in {\cal N}$ where $\phi = \frac{\sqrt{(\alpha-K)^2 + 4 \alpha K N} + \alpha - K}{2 K}$, the competitive ratio ${\tt CR}({\cal A}^{{\rm itc.}K}_{\rm thb}[{\textstyle\frac{\rho}{(1-\rho)\phi}}]) = \frac{\sqrt{(\alpha-K)^2 + 4 \alpha K N} + \alpha + K}{2 K}$.
\end{customthm}

%{\bf Remark:} Similar to the setting of no intercommunication, the competitive ratio of ${\cal A}^{{\rm itc.}K}_{\rm thb}$ also diminishes with $K$. %The intuition is that the optimal offline solution needs to produce more report messages per each event, which incur a communication cost closer to that of ${\cal A}^{{\rm itc.}K}_{\rm thb}$.

\section{Partial Intercommunication}  \label{sec:net}

This section considers non-ideal intercommunication, modelled by a communication graph. A communication graph is defined by ${\cal G} = ({\cal N}, {\cal E})$, where each edge $(i, i') \in {\cal E}$ represents that a pair of systems that can engage in full intercommunication (including overhearing) with each other.

We consider a general process that consists of two options:
\begin{enumerate}

\item {\em Overhear-and-withhold}: When a system overhears the report messages from neighboring systems, it will not generate a report message for those events that have been reported $K$ times. It also adjusts the accumulative latency penalty for the unreported events.  

\item {\em Overhear-and-forward}: When a system overhears the report messages from neighboring systems, it will replicate the report messages, combining with its on-going report message to its neighbors. It is assumed that there is collision-free intercommunication with the neighbors. 

\end{enumerate}

%\vspace{-10pt}
\begin{figure}[htb!] 
\centering
\includegraphics[scale=0.55]{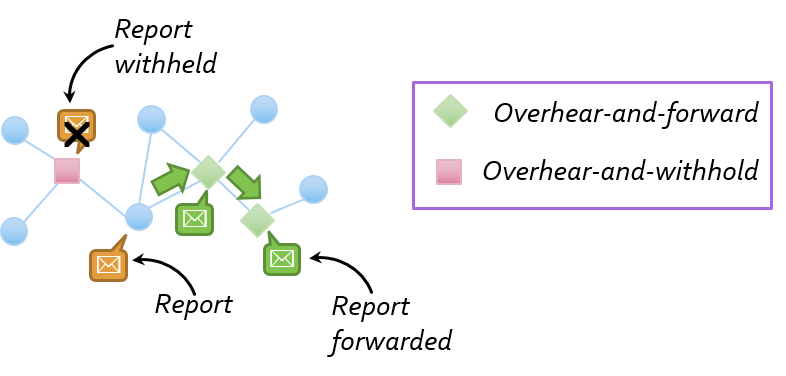}  
%\vspace{-25pt}
\caption{\small   An illustration for overhear-and-withhold and overhear-and-forward operations. \label{fig:net1}}   
%\vspace{-10pt}
\end{figure}

We partition the systems into the two sets. Let ${\cal N}_{\rm w} \subseteq {\cal N}$ be the set of overhear-and-withhold nodes, and ${\cal N}_{\rm f} = {\cal N} \backslash {\cal N}_{\rm w}$ be the set of overhear-and-forward nodes. Threshold-based algorithm ${\cal A}^{{\rm net.}K}_{\rm thb}$ (Algorithm~\ref{alg:net}) performs operations according to the role of each node.

\begin{algorithm}
\caption{\small   ${\cal A}^{{\rm net.}K}_{\rm thb}[\theta, t_{\rm now}, (\sigma_i^j)_{j:{\tt t}^j\le t_{\rm now}}, ({z}_{i'}^j(t_{\rm now}))_{i' \in{\cal N}: (i, i') \in {\cal E}}]$} \label{alg:net}
{\small  \begin{algorithmic}[1] 
\Require Threshold $\theta$; current time $t_{\rm now}$; known measurements $(\sigma_i^j)_{j:{\tt t}^j\le t_{\rm now}}$; indicators of reported events ${z}_{i'}^j(t_{\rm now})$ from neighbors.
\Statex \hspace{-15pt} {\bf Global Init.:} $x_k(i,j) \leftarrow 0$ for all $j \in {\cal M}, k \leftarrow 0, r_{i,k} \leftarrow 0$
\vspace{3pt} \hrule \vspace{3pt}
\Statex \Comment{{\em Count num. of reports from neighbors}}
\For{each $j$ such that $r_{i,k} < {\tt t}^j \le t_{\rm now}$}
\State ${\rm cnt}_j \leftarrow \sum_{i' : (i,i')\in{\cal E}} {z}_{i'}^j(t_{\rm now})$ 
\EndFor
\Statex \Comment{{\em Compute latency penalty for events reported $<$ K times}}
\State ${\tt lat} \leftarrow \displaystyle \sum_{j \in {\cal M} : r_{i,k} < {\tt t}^j \le t_{\rm now} \wedge {\rm cnt}_j < K} {\cal L}(\sigma_i^j, t_{\rm now})$
\Statex \Comment{{\em Compute communication cost for events reported $<$ K times}}
\State ${\tt com} \leftarrow {\cal C}\Big(\{ \sigma_i^j:  j \in {\cal M}, r_{i,k} < {\tt t}^j \le t_{\rm now} \wedge {\rm cnt}_j < K \}\Big)$
\Statex
\If {$\frac{{\tt lat}}{{\tt com}} \ge \theta$} \Comment{{\em Condition on threshold}}
\If {$i \in {\cal N}_{\rm f}$} \Comment{{\em Overhear-and-forward}}
\State $k \leftarrow k + 1$\Comment{{\em Next report message}}
\State $x_k(i,j) \leftarrow 1$ for all observed $j$ or if $z^j_{i'}(t_{\rm now})=1$ 
\State $z^j_i(t_{\rm now}+\epsilon) \leftarrow 1$ for all reported $j$ or if $z^j_{i'}(t_{\rm now})=1$  
\State $r_{i,k} \leftarrow t_{\rm now}$ 
\ElsIf {$i \in {\cal N}_{\rm w}$} \Comment{{\em Overhear-and-withhold}}
\State $k \leftarrow k + 1$\Comment{{\em Next report message}}
\State $x_k(i,j) \leftarrow 1$ for all observed event $j$ 
\State $z^j_i(t_{\rm now}+\epsilon) \leftarrow 1$ for all reported event $j$, and any $\epsilon$
\State $r_{i,k} \leftarrow t_{\rm now}$ 
\EndIf
\EndIf
\end{algorithmic}}
\end{algorithm}

Let ${\tt x}$ be the maximum number of systems that can produce simultaneous report messages in ${\cal G}$. That is,
\begin{equation} \label{eqn:x} \hspace{-5pt}
 {\tt x} \triangleq \max_{{\cal X} \subset {\cal N}_{\rm w} }\Big|\{{\cal N}_{\rm f} \cup {\cal X}  \mid i \in {\cal X} \Rightarrow \nexists i' \in  {\cal N}_{\rm f} \cup {\cal X} \mbox{\ and\ } (i, i') \in {\cal E} \}\Big|  \notag
\end{equation}
Namely, ${\tt x}$ is the number of nodes in ${\cal N}_{\rm f}$, and the maximum number of nodes in ${\cal N}_{\rm w}$ that are not the neighbors of each other or with a node in ${\cal N}_{\rm f}$.
Here are a few examples of ${\tt x}$:
\begin{itemize}

\item[(${\tt NC}$)] For no intercommunication, we set ${\cal N}_{\rm w} = {\cal N}$ and ${\cal E} = \varnothing$, then ${\tt x} = N$.

\item[(${\tt FC}$)] For full intercommunication, we set ${\cal N}_{\rm w} = {\cal N}$ and ${\cal E} = \{ (i, i')\}_{i, i' \in {\cal N}}$, then ${\tt x} = 1$.

\item[(${\tt N1}$)] All nodes carry out overhear-and-withhold: ${\cal N}_{\rm w} = {\cal N}$. Let the maximum independent set of ${\cal G}$ by ${\tt MIS}({\cal G})$. Then ${\tt x} = |{\tt MIS}({\cal G})|$.

\item[(${\tt N2}$)] The nodes that belong to a minimum connected dominating set (${\tt CDS}$) carry out overhear-and-forward, while the rest of nodes carry out overhear-and-withhold. We assume that ${\cal G}$ is connected. Let the minimum connected dominating set by ${\tt CDS}({\cal G})$. Let ${\cal N}_{\rm f} = {\tt CDS}({\cal G})$, ${\cal N}_{\rm w} = {\cal N} \backslash {\tt CDS}({\cal G})$. Then ${\tt x} = |{\tt CDS}({\cal G})|$.

\end{itemize}

\smallskip

\begin{customthm}{5} \label{thm:net}
Given a communication graph ${\cal G}$, with ${\cal N}_{\rm w}$ and ${\cal N}_{\rm f}$, Setting threshold $\theta = {\textstyle\frac{\rho}{(1-\rho)\phi}}$ for every system $i \in {\cal N}$, where 
\begin{equation}
\phi = \frac{\sqrt{(\alpha {\rm x} -K)^2 + 4 \alpha K N} + \alpha {\rm x} - K}{2 K}
\end{equation}
the competitive ratio of ${\cal A}^{{\rm net.}K}_{\rm thb}$ is 
\begin{equation}
{\tt CR}({\cal A}^{{\rm net.}K}_{\rm thb}[{\textstyle\frac{\rho}{(1-\rho)\phi}}]) = \frac{\sqrt{(\alpha {\rm x} - K)^2 + 4 \alpha K N} + \alpha {\rm x} + K}{2 K} 
 \label{eq:netcr}
\end{equation}
\end{customthm}

{\bf Remark:} For no intercommunication (${\rm x} = N$), it follows that the competitive ratio becomes ${\tt CR}({\cal A}^{{\rm net.}K}_{\rm thb}[{\textstyle\frac{\rho}{(1-\rho)\phi}}]) =\frac{\alpha N}{K}+1$ $(= {\tt CR}({\cal A}_{\rm thb}[{\textstyle\frac{K \rho}{\alpha N(1-\rho)}}]))$.
For full intercommunication (${\rm x} = 1$), it follows that the competitive ratio becomes ${\tt CR}({\cal A}^{{\rm net.}K}_{\rm thb}[{\textstyle\frac{\rho}{(1-\rho)\phi}}]) = \frac{\sqrt{(\alpha-K)^2 + 4 \alpha K N} + \alpha + K}{2 K}$ $(={\tt CR}({\cal A}^{{\rm itc.}K}_{\rm thb}[{\textstyle\frac{\rho}{(1-\rho)\phi}}]))$. 
Thus, ${\cal A}^{{\rm net.}K}_{\rm thb}[{\textstyle\frac{\rho}{(1-\rho)\phi}}]$ provides a continuous transition between the two cases.

\section{Lower Bounds of Competitive Ratios} \label{sec:lb}

This section shows that the threshold-based algorithms (${\cal A}_{\rm thb}$ and ${\cal A}^{{\rm itc.}K}_{\rm thb}$) achieve the best competitive ratios in terms of order of magnitude among all deterministic distributed online algorithms in the respective settings. It suffices to consider only {\rm $1$-DIA} for establishing the lower bounds of competitive ratios.

In practice, intercommunication consumes at least one transmission but still may require an extra transmission for reporting. We assume that {\em any intercommunication among systems is as costly as transmitting a report message}. 

\smallskip

\begin{customthm}{6}\label{thm:woc-lb}
Consider {\rm $1$-DIA}. Suppose $\rho=\frac{1}{2}$ and there is one system having a communication cost per report as 1, while other $(N-1)$ systems having a communication cost per report as $\alpha \ge 1$. There exists no deterministic distributed online algorithm with no intercommunication to achieve a competitive ratio lower than $\alpha(N-1)+\frac{3}{2}$.
\end{customthm}

\smallskip

\begin{customthm}{7} \label{thm:whc-lb}
Consider {\rm $1$-DIA} and $\rho=\frac{1}{2}$. There exists no deterministic distributed online algorithm with collision-free intercommunication to achieve a competitive ratio lower than $\frac{\sqrt{N}}{4}$.
\end{customthm}

\smallskip

{\bf Remark:} Theorems~\ref{thm:woc-lb}-\ref{thm:whc-lb} show that the competitive ratios for threshold-based algorithms in Theorem~\ref{thm:woc.alg} and Theorem~\ref{thm:whc.alg} (i.e., $O(N)$ and $O(\sqrt{N})$ respectively) are optimal in order of magnitude.
\section{Simulation Studies} \label{sec:sim}

We provided analysis on the competitive ratios of our algorithms in the previous sections, which are the worst-case guarantees. In this section, we evaluate the empirical {\em average-case} ratios by simulations. 
%We employ relatively large network topologies generated by random unit disk graphs (up to 300 systems). 
We compare the average-case ratios with our theoretical competitive ratios, and observe that our algorithms perform relatively well in several scenarios which are far below the theoretical worst-case values.

\begin{figure*}[!ht] %\vspace{-10pt} 
\hspace{-15pt}
\begin{subfigure}[h]{0.25\textwidth}
	\includegraphics[scale=.68]{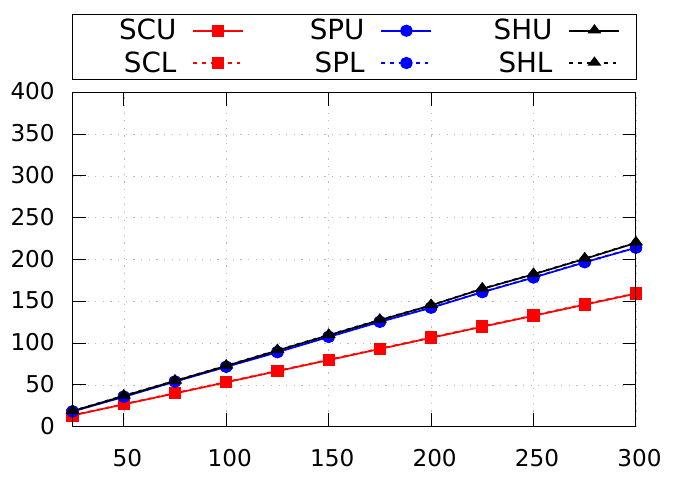}
	\vspace{-10pt}
	\caption{\small No intercommunication}
	\label{fig:nocom-small}
\end{subfigure}
\begin{subfigure}[h]{0.25\textwidth}
	\includegraphics[scale=.68]{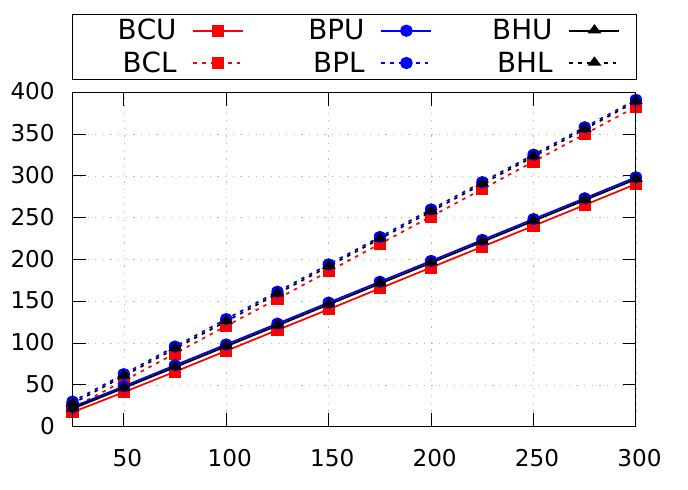}
	\vspace{-10pt}
	\caption{\small No intercommunication}
	\label{fig:nocom-big}
\end{subfigure}
\begin{subfigure}[h]{0.25\textwidth}
	\includegraphics[scale=.68]{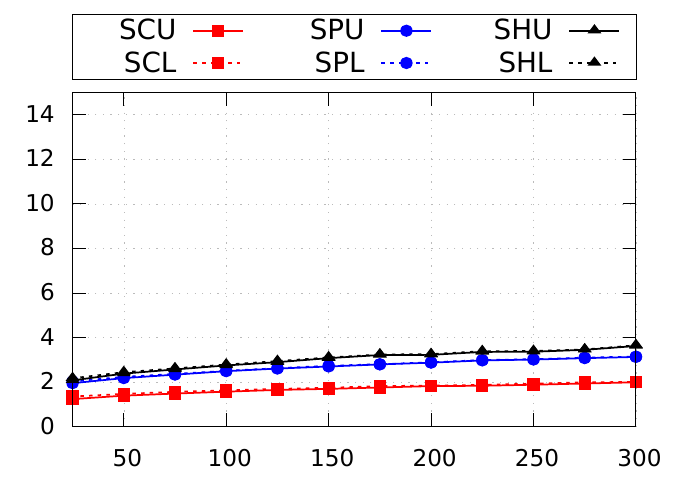}
	\vspace{-10pt}
	\caption{\small Full intercommunication}
	\label{fig:fullcom-small}
\end{subfigure}
\begin{subfigure}[h]{0.25\textwidth}
	\includegraphics[scale=.68]{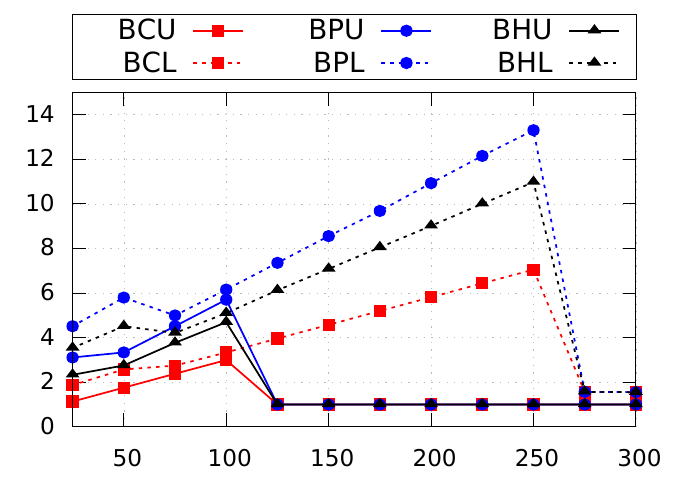}
	\vspace{-10pt}
	\caption{\small Full intercommunication}
	\label{fig:fullcom-big}
\end{subfigure}
\caption{\small Different event arrival models for the settings of no intercommunication and full intercommunication. The $y$-axis is the empirical ratio, whereas the $x$-axis is the number of systems $N$.} \label{fig:sim1}
%\vspace{-10pt}
\end{figure*}

%%%%%%%%%%%%%%%%%%%%%%%%%%%%%%%%%%%%%%
%%%%%%%%%%%%%%% rho %%%%%%%%%%%%%%%%%%
%%%%%%%%%%%%%%%%%%%%%%%%%%%%%%%%%%%%%%
\begin{figure}[!ht] %\vspace{-10pt} 
	\hspace{-15pt}
	\begin{subfigure}[h]{0.25\textwidth}
		\includegraphics[scale=.68]{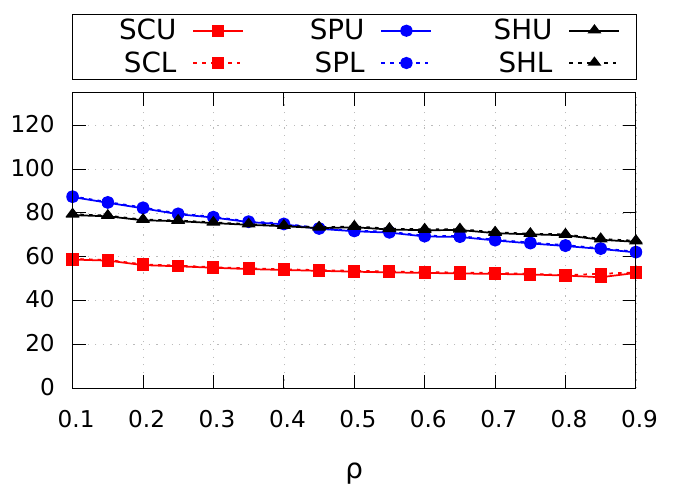}
		\vspace{-10pt}
		\caption{\small No intercommunication}
		\label{fig:rnocom-small}
	\end{subfigure}
	\begin{subfigure}[h]{0.25\textwidth}
		\includegraphics[scale=.68]{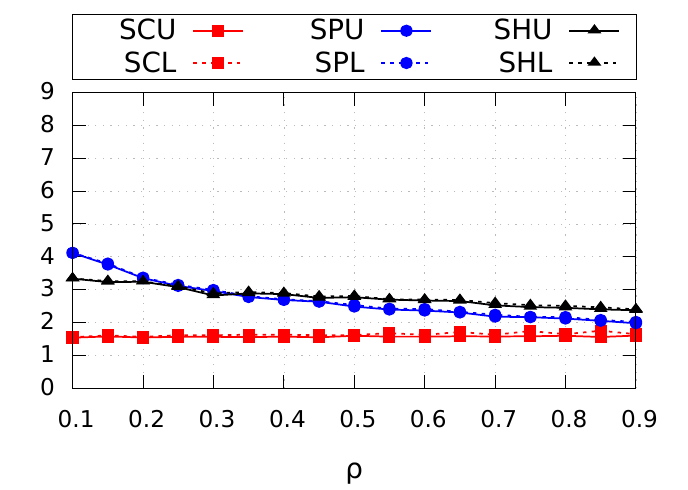}
		\vspace{-10pt}
		\caption{\small Full intercommunication}
		\label{fig:rcom-small}
	\end{subfigure}
	\caption{\small The $x$-axis corresponds to the different values of  $\rho$, whereas  the $y$-axis is the empirical ratio.} \label{fig:simr}
	%\vspace{-10pt}
\end{figure}
%%%%%%%%%%%%%%%%%%%%%%%%%%%%%%%%%%%%%%
%%%%%%%%%%%%%%%%%%%%%%%%%%%%%%%%%%%%%%
%%%%%%%%%%%%%%%%%%%%%%%%%%%%%%%%%%%%%%

%%%%%%%%%%%%%%%%%%%%%%%%%%%%%%%%%%%%%%
%%%%%%%%%%%%%%% theta %%%%%%%%%%%%%%%%%%
%%%%%%%%%%%%%%%%%%%%%%%%%%%%%%%%%%%%%%
\begin{figure}[!ht] %\vspace{-10pt} 
	\hspace{-15pt}
	\begin{subfigure}[h]{0.25\textwidth}
		\includegraphics[scale=.68]{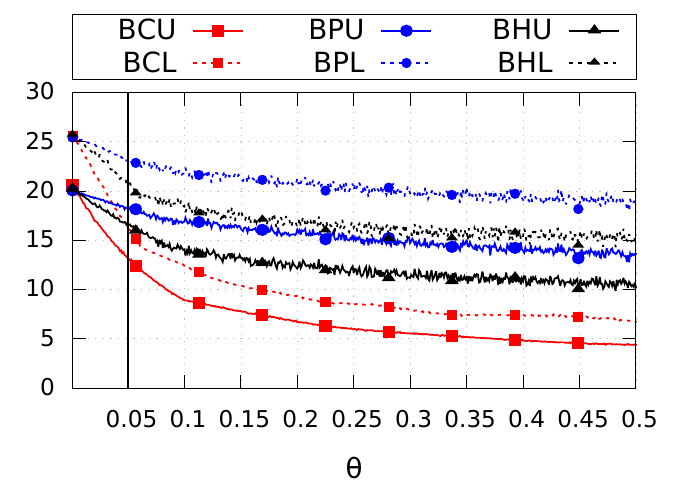}
		\vspace{-10pt}
		\caption{\small No intercommunication}
		\label{fig:theta-nocom-big}
	\end{subfigure}	
	\begin{subfigure}[h]{0.25\textwidth}
		\includegraphics[scale=.68]{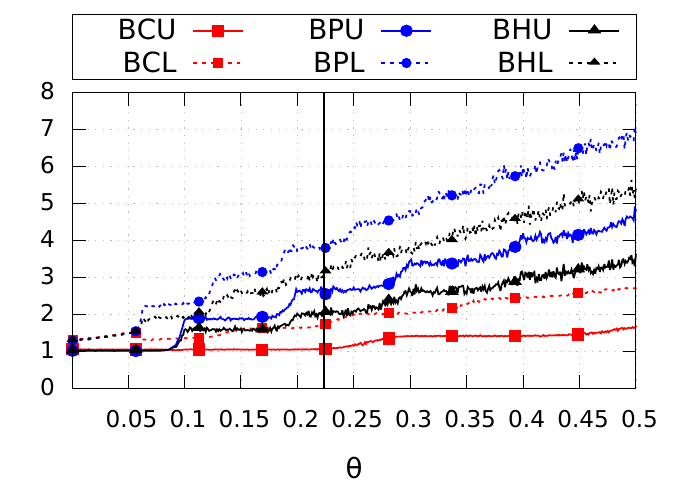}
		\vspace{-10pt}
		\caption{\small Full intercommunication}
		\label{fig:theta-com-big}
	\end{subfigure}
	\caption{\small The $x$-axis corresponds to the different values of  $\theta$, whereas  the $y$-axis is the empirical ratio. The solid vertical lines correspond to the values of $\theta$ derived by our theorems.} \label{fig:simt}
	%\vspace{-10pt}
\end{figure}
%%%%%%%%%%%%%%%%%%%%%%%%%%%%%%%%%%%%%%
%%%%%%%%%%%%%%%%%%%%%%%%%%%%%%%%%%%%%%
%%%%%%%%%%%%%%%%%%%%%%%%%%%%%%%%%%%%%%

%%%%%%%%%%%%%%%%%%%%%%%%%%%%%%%%%%%%%%
%%%%%%%%%%%%%%% worst %%%%%%%%%%%%%%%%
%%%%%%%%%%%%%%%%%%%%%%%%%%%%%%%%%%%%%%
\begin{figure}[!ht] %\vspace{-10pt} 
	\hspace{-15pt}
	\begin{subfigure}[h]{0.25\textwidth}
		\includegraphics[scale=.68]{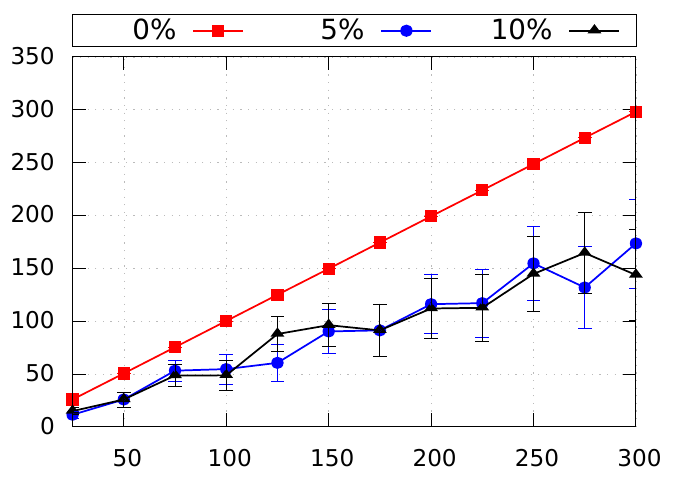}
		\vspace{-10pt}
		\caption{\small No intercommunication}
		\label{fig:worst-nocom}
	\end{subfigure}	
	\begin{subfigure}[h]{0.25\textwidth}
		\includegraphics[scale=.68]{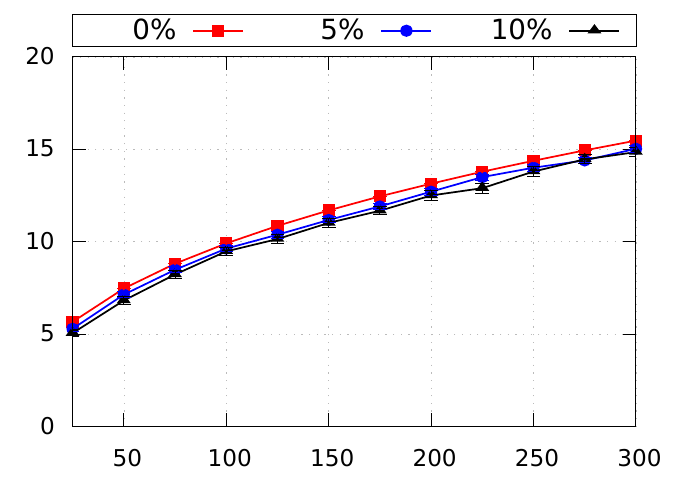}
		\vspace{-10pt}
		\caption{\small Full intercommunication}
		\label{fig:worst-com}
	\end{subfigure}
	
	\caption{\small Simulations of randomly perturbed worst-case inputs, where the measurements are perturbed by a random percent. The $x$-axis corresponds to the different values of  $N$, whereas  the $y$-axis is the empirical ratio.}
	 \label{fig:worst}
	%\vspace{-10pt}
\end{figure}
%%%%%%%%%%%%%%%%%%%%%%%%%%%%%%%%%%%%%%
%%%%%%%%%%%%%%%%%%%%%%%%%%%%%%%%%%%%%%
%%%%%%%%%%%%%%%%%%%%%%%%%%%%%%%%%%%%%%

%%%%%%%%%%%%%%%%%%%%%%%%%%%%%%%%%%%%%%
%%%%%%%%%%%%%%%%% net %%%%%%%%%%%%%%%%
%%%%%%%%%%%%%%%%%%%%%%%%%%%%%%%%%%%%%%
\begin{figure*}[!ht] %\vspace{-10pt} 
  \hspace{-20pt}    
\begin{subfigure}[h]{0.2\textwidth}
\centering
\includegraphics[scale=0.2,trim = 1cm 5.7cm 1cm 3.0cm, clip=true]{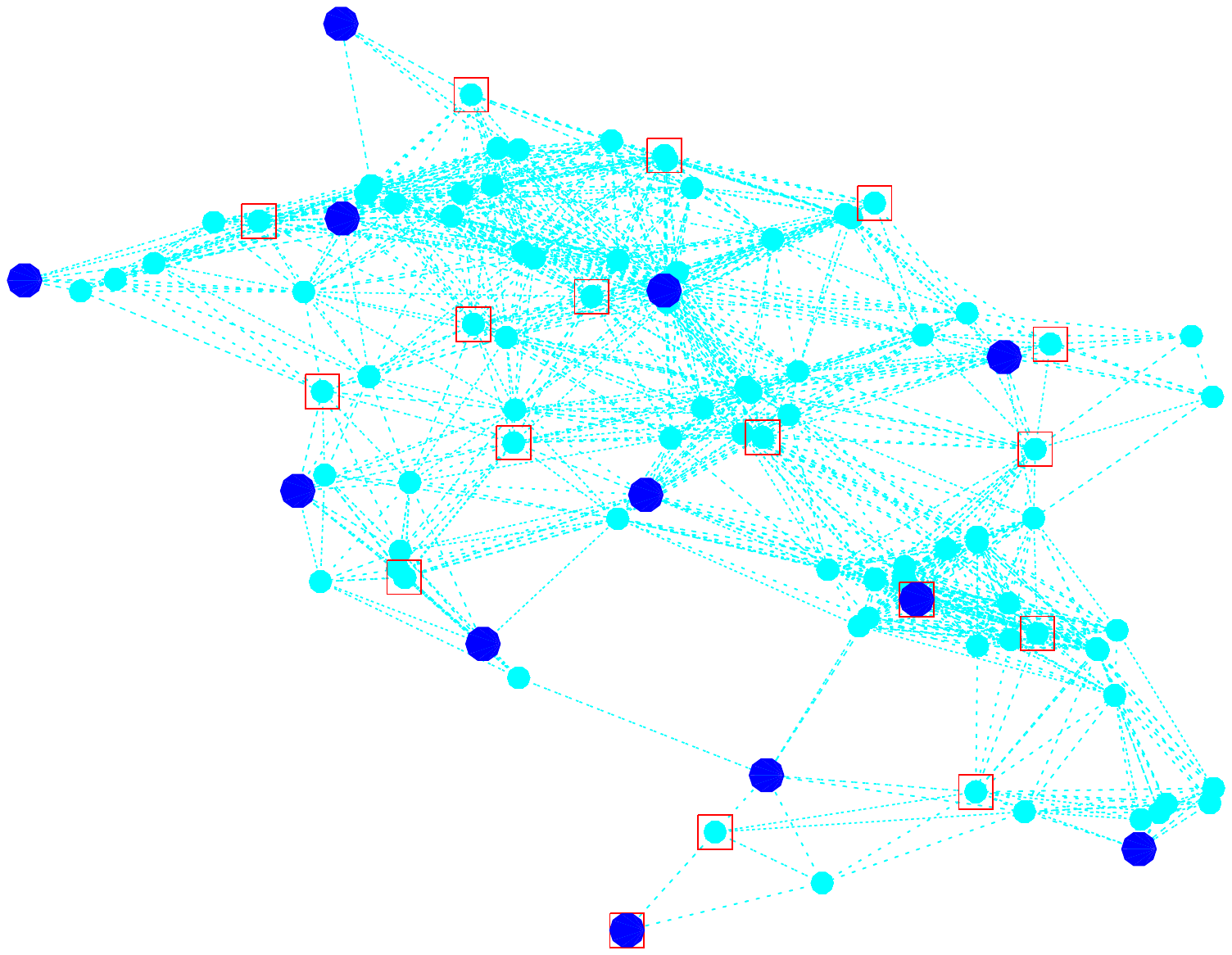}
\caption{\small A random UDG}
\label{fig:rnd-toplogy}
\end{subfigure}        
  \begin{subfigure}[h]{0.2\textwidth}
	\includegraphics[scale=.75]{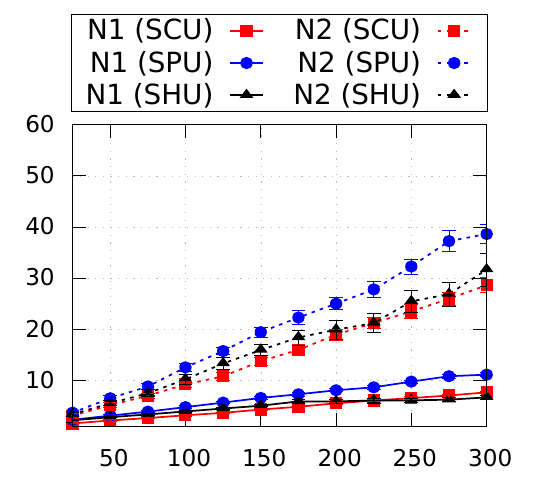}
	\vspace{-10pt}
	\caption{\small   Unity cost}
	\label{fig:sim1-SCU-SPU-SHU}
\end{subfigure}
\begin{subfigure}[h]{0.2\textwidth}
	\includegraphics[scale=.75]{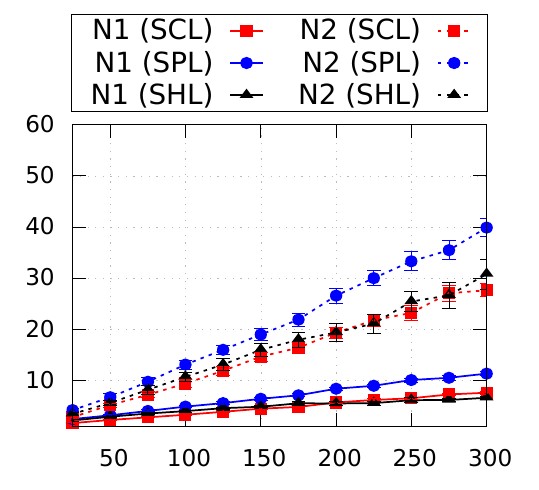}
	\vspace{-10pt}
	\caption{\small Log cost}
	\label{fig:sim1-SCL-SPL-SHL}
\end{subfigure} %\vspace{-10pt}
\begin{subfigure}[h]{0.2\textwidth}
	\includegraphics[scale=.75]{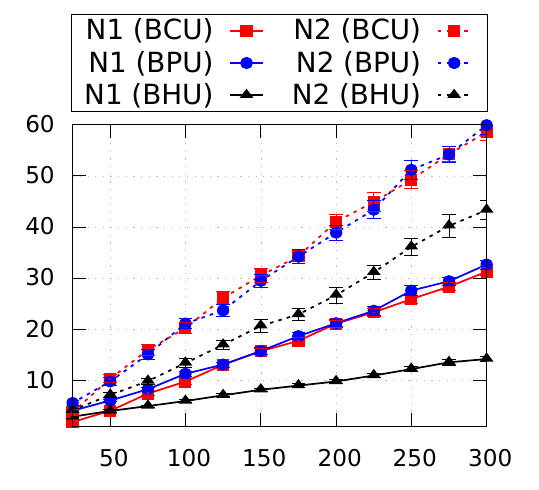}
	\vspace{-10pt}
	\caption{\small Unity cost}
	\label{fig:sim1-BCU-BPU-BHU}
\end{subfigure}
\begin{subfigure}[h]{0.2\textwidth}
   \includegraphics[scale=0.75]{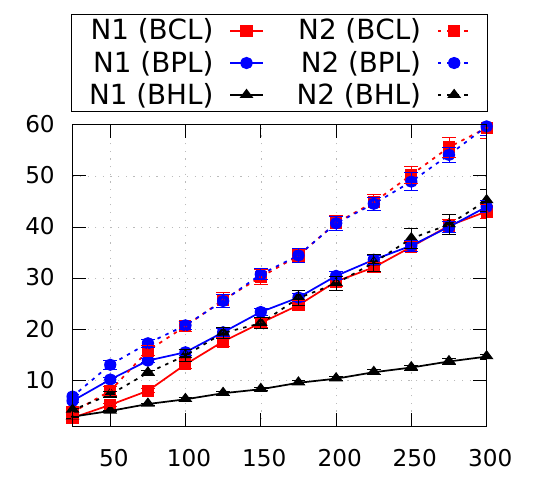}
	 \vspace{-10pt}
	\caption{\small Log cost}
	\label{fig:sim1-BCL-BPL-BHL}
\end{subfigure} %\vspace{-10pt}
   \caption{\small (a) An example of a random Unit Disk Graph (UDG) of $100$ nodes that is generated randomly in the simulation. The red squares represent a ${\tt CDS}$, while the blue circles belong to ${\tt MIS}$. (b-e) The figures show different scenarios for the setting of partial intercommunication. The $y$-axis is the empirical ratio, whereas the $x$-axis is the number of systems $N$. } \label{fig:net}
%\vspace{-10pt}		
\hspace{-20pt} 
\end{figure*}
%%%%%%%%%%%%%%%%%%%%%%%%%%%%%%%%%%%%%%
%%%%%%%%%%%%%%%%%%%%%%%%%%%%%%%%%%%%%%
%%%%%%%%%%%%%%%%%%%%%%%%%%%%%%%%%%%%%%

\subsection{Simulation Set-up}

%The software environment for simulation is GNU Octave. 
The simulations were performed up to 300 systems, and each has 50 runs. For the partial intercommunication, we generate a random connected unit disk graphs with a fixed average degree ($\approx 18$) for each run. 
%We use a $6$-approximation and $3$-approximation algorithms for {\tt {\tt CDS}} and {\tt {\tt MIS}}, respectively in each random graph. 
Fig.~\ref{fig:rnd-toplogy} shows an example of a random network of $100$ systems in the simulation.

The measurement values $(w^i_j)$ are uniformly distributed generated by the following two ranges:
\begin{enumerate}
\item {\em Big events} (B): $w^i_j \in [0,1]$ for all $i, j$.
\item {\em Small events} (S): $w^i_j \in [0,\frac{1}{N}]$ for all $i, j$.
%\item {\em Local events} (L):  $w^i_j \in [0,1]$ for only one random system $i$ per event $j$.
\end{enumerate}
The appearance times $({\tt t}^j)$  are generated by the following three random arrival models: 
\begin{enumerate}
\item {\em Constant-interval arrivals} (C): The inter-arrival time is a constant. %The experiments for this model span through $100$ time slots.
\item {\em Poisson arrivals} (P): The inter-arrival time follows an exponential distribution with mean of $20$ time slots. %Since each time slot duration is $1/100$, the rate indicates an average of $50$ time slots between subsequent events. The total number of events is fixed to $100$ for each run while the total time slots is kept random.
\item {\em Heavy-tailed arrivals} (H): The inter-arrival time follow Weibull distribution with shape parameter $0.5$, and mean of $10$ time slots.
\end{enumerate}
The latency penalty function is chosen to be linear, while the following two communication cost functions are considered:
\begin{enumerate}
\item {\em Unity cost} (U): The communication cost is always $1$, ${\cal C}(\sigma_{i,k}) = 1$.
\item {\em Logarithmic cost} (L): The communication cost is a logarithm function of local total measurement value, ${\cal C}(\sigma_{i,k}) =  \log\Big(2+\sum_{j\in{\cal M}: {\sigma}_i^j \in {\sigma}_{i,k}} w_i^j \Big)$.
\end{enumerate}
In the following, a simulation scenario is represented by the acronyms. For instance, BHL represents the scenario with big events, heavy-tailed arrivals and logarithmic cost.

\subsection{Observations and Discussion}
\begin{figure}[!ht] %\vspace{-10pt} 
         \begin{subfigure}[h]{0.23\textwidth}
                \includegraphics[scale=.68]{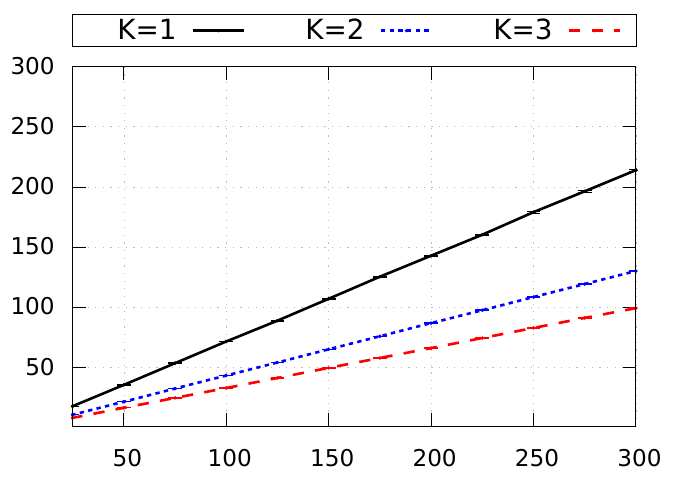}
                \vspace{-10pt}
                \caption{\small No comm. (SPU)}
                \label{fig:k-nocom}
        \end{subfigure}
         \begin{subfigure}[h]{0.23\textwidth}
                \includegraphics[scale=.68]{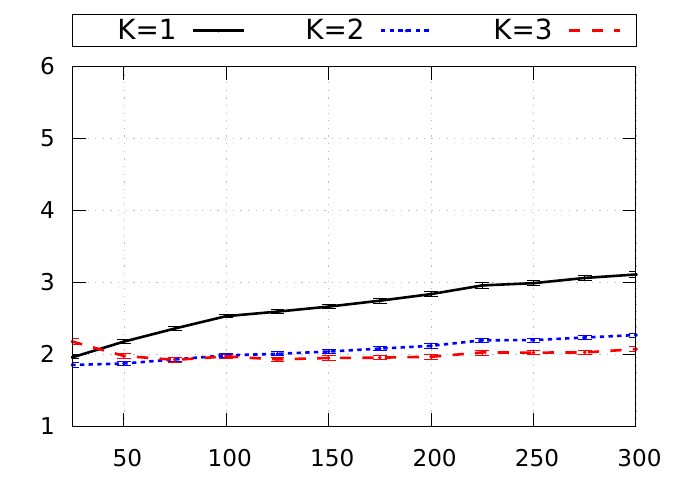}
                \vspace{-10pt}
                \caption{\small Full comm. (SPU)}
                \label{fig:k-fullcom}
        \end{subfigure}
        \caption{\small The effect of different values of $K$ on the empirical ratio for the settings of no intercommunication and full intercommunication for scenario SPU.} \label{fig:k-com}
 \vspace{-10pt}
\end{figure}

We present the simulation results for each scenario, and discuss the observations.
\begin{enumerate}

\item {\bf No Intercommunication}:
Figs.~\ref{fig:nocom-small}-\ref{fig:nocom-big} show the empirical ratios for different scenarios. Overall, we observe linear trends in the empirical ratios as $N$ increases, which are below the respective theoretical competitive ratios. Note that constant-interval arrivals (C) give consistently better empirical ratios, followed by Poisson arrivals (P) and heavy-tailed arrivals (H). However, when using big events (Fig.~\ref{fig:nocom-big}), the difference in empirical ratio is much smaller. We also observe that logarithmic cost (L) results in higher empirical ratios than unity cost which is justified by the additional factor $\alpha$ in Theorem~\ref{thm:woc.alg}. %Moreover, we evaluate the empirical ratios in Figs.~\ref{fig:norm-nocom-small}-\ref{fig:norm-nocom-big}, considering the latency normalized by a factor $\frac{1}{N}$, which show superior performance.

We present a sensitivity analysis for various values of $\rho$ and $\theta$. We set the number of systems as 100.  Fig.~\ref{fig:rnocom-small} shows that higher value of $\rho$ achieves the better (i.e., lower) empirical ratio. Fig.~\ref{fig:theta-nocom-big} illustrates the effect of using different values of $\theta$ rather than the one derived by our theorems, which is shown by a vertical line. The higher value $\theta$ achieves the better empirical ratio. 

Fig.\ref{fig:worst-nocom} shows the empirical ratios of randomly perturbed worst-case input from Theorem~\ref{thm:woc-lb}, where the measurements are perturbed by a random percent. Slightly perturbed inputs significantly reduce the empirical ratio.

Fig.~\ref{fig:k-nocom} shows the empirical ratios for the $K$-report problem when $K=1,2,3$ for scenario SPU. A larger $K$ yields a lower empirical ratio, as shown in Theorem~\ref{thm:woc.alg.k}.

\item {\bf Full Intercommunication}: 
Figs.~\ref{fig:fullcom-small}-\ref{fig:fullcom-big} show the empirical ratio for different scenarios. Overall, we observe a sub-linear trend in the empirical ratio as $N$ increases, which are below the theoretical competitive ratios. For small events, the empirical ratios increase smoothly. However, for big events, the empirical ratios drop to 1, as $N$ reaches a certain value. The intuition for such a drop is that when $N$ is large, the total latency penalty exceeds the communication cost for almost all events $j$ at every time slot, therefore an optimal offline solution generates a report message at every time slot, which coincides with the threshold-based algorithm. 

Fig.~\ref{fig:rcom-small} shows that different values of  $\rho$ has small effect on the empirical ratio for small events. Fig.~\ref{fig:theta-com-big} illustrates the effect of using different values of  $\theta$. Increasing $\theta$ beyond that is derived by our theorem increases the empirical ratio. Fig.\ref{fig:worst-com} shows the empirical ratios of randomly perturbed worst-case input from Theorem~\ref{thm:whc-lb}.

Fig.~\ref{fig:k-fullcom} shows the empirical ratios for  the $K$-report problem when $K=1,2,3$ for scenario SPU. A larger $K$ yields a lower empirical ratio.

\item {\bf Partial Intercommunication}: 
Figs.~\ref{fig:sim1-SCU-SPU-SHU}-\ref{fig:sim1-BCL-BPL-BHL} show the empirical ratios for partial intercommunication using settings of ${\tt N1}$ and ${\tt N2}$ in Sec.~\ref{sec:net}. We observe that ${\tt N1}$ obtains smaller empirical ratios than ${\tt N2}$ for all scenarios. 
We used approximation algorithms to obtain ${\tt MIS}$ and ${\tt CDS}$.  
In most scenarios we observe that ${\tt CDS}$ is slightly larger than ${\tt MIS}$ (see Fig.~\ref{fig:rnd-toplogy} for an example), which creates a larger empirical ratio for ${\tt N2}$.
We also observe that small events give smaller empirical ratios than large events in all scenarios. Moreover, logarithmic cost gives larger empirical ratios as expected.

\end{enumerate}

%%% nocom/ different k

%%\vspace{-5pt}
\section{Practical Sensor Network Testbed} \label{sec:impl}

As a concrete demonstration for our distributed online algorithms, we implemented our threshold-based algorithms in a practical sensor network testbed for applications such as intrusion detection and air-conditioning control. For each application, we deploy a network of wireless sensor nodes equipped with appropriate sensors for detecting the relevant signals from the environment, which then transmit report messages to a base station. 

%
%%\vspace{-5pt}
\subsection{Hardware Set-up}

\begin{figure}[htb!] 
  \centering	  
  \includegraphics[scale = 0.3]{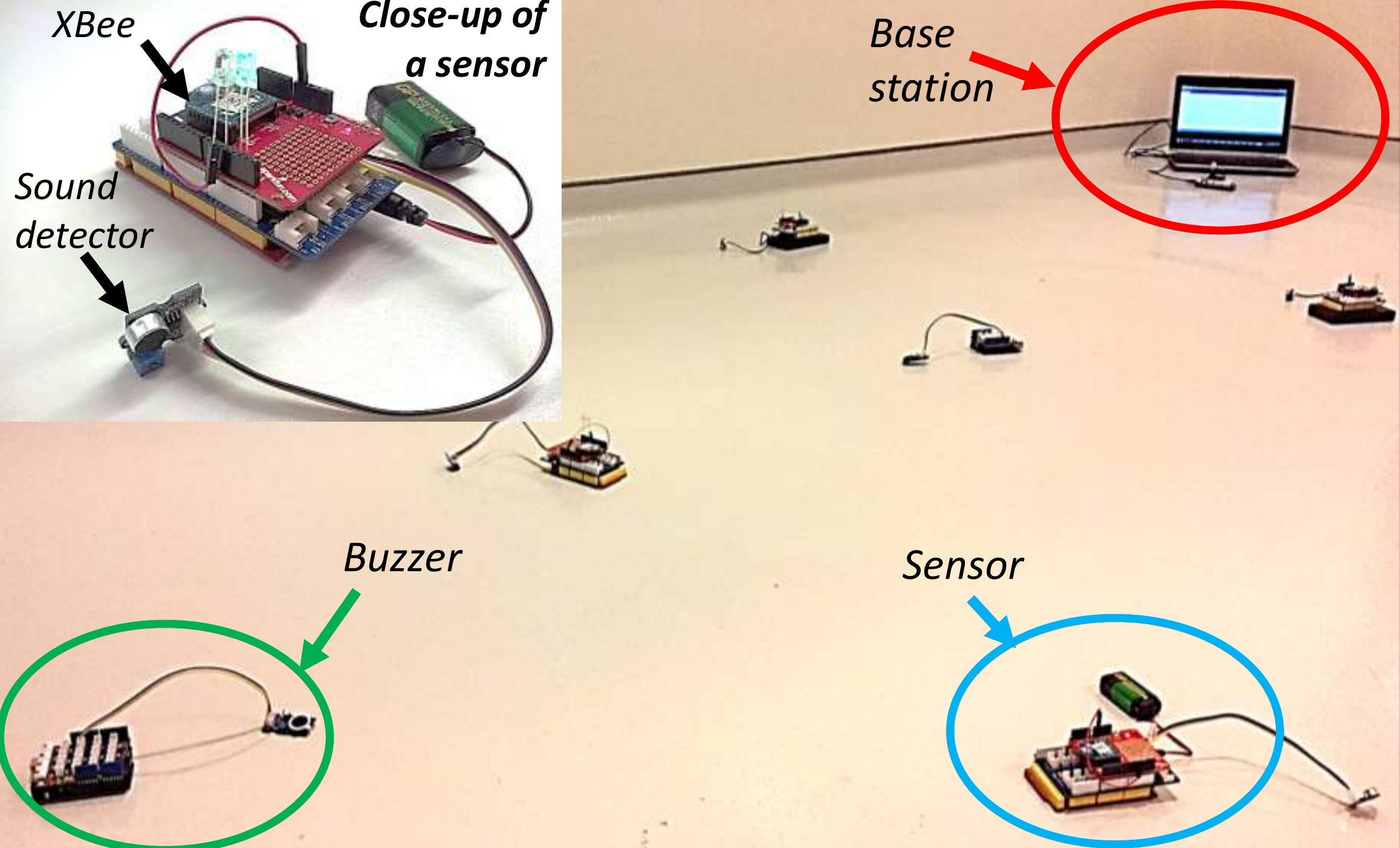} 
  %%\vspace{-5pt}
  \caption{\small   A snapshot of the sensor network testbed.}
  \label{fig:expsetup}
\end{figure}

The sensor network testbed is deployed in an office building (see Fig.~\ref{fig:expsetup}). The testbed consists of:

\begin{enumerate}
\item A network of wireless sensor nodes programmed with our threshold-based algorithms. We use several types of sensors to monitor various events depending on the applications. For example, the sound sensor is an XBee-enabled Arduino board equipped with a sound detector while the remaining sensors are TinyOS based. Each sensor can be set with broadcast mode, which can overhear other sensors' transmissions.    
%\vspace{-6pt}
%\item One XBee-enabled Arduino that is connected to a laptop computer and acts as a base station for other sensors. It records data received from wireless sensors for offline analysis.
%\vspace{-6pt}
\item Several TinyOS-based base stations that are also connected to a laptop computer and record data from received from TinyOS based sensor nodes.
%\vspace{-6pt}
\item Several Arduino boards equipped with buzzers and lights that can act as random fluctuating sound and light sources, and generate measurement for the sound sensors and light sensors. 
%\vspace{-6pt}
\item A room heater controlled by Arduino that acts as random fluctuating heat source to generate measurements for temperature sensors.
%\vspace{-6pt}
\item A hardware module that monitors the energy consumptions of wireless sensors.
\end{enumerate}

%%\vspace{-5pt}

To implement collision-free intercommunication, we use a simple priority based mechanism. Each sensor node is assigned priority order in advanced, which determines for how long the sensor must wait (in multiple of 10ms) before transmitting data over the radio channel. 
%This ensures that sensors with high priority always transmit first. 
The configurations for each sensor are remotely controllable from the base station.

%%\vspace{-5pt}
\subsection{Empirical Results and Discussion}

We present the empirical evaluation of the competitive ratios of our threshold-based algorithms. We conducted the experiments for 1-report problem with (1) no intercommunication, where the wireless broadcast mode is off, and (2) full intercommunication, where the wireless broadcast mode is on. 

Each wireless sensor is programmed to take measurements at appropriate intervals. Whenever the accumulative latency as a function of measured signal exceeds the specified threshold in our algorithms, the wireless sensor will transmit a notification to the base station via ZigBee wireless link. In case the broadcast mode is turned on and the accumulative latency exceeds the threshold for multiple sensors simultaneously, only the sensor with highest priority will transmit the report. 

We next describe the empirical observations. For each type of sensor, we depict a sample trace of the measured signals and plot the results in terms of competitive ratios in Fig.~\ref{fig:sensor_res}. To compute the competitive ratios, we divide the total cost by the respective optimal offline cost.

\smallskip

%%\vspace{-5pt}
\subsubsection{Sensor Data}

\begin{figure*}[htb!] 
    \hspace{-0pt}       \begin{subfigure}[h] {0.20\textwidth}
              \includegraphics[scale = 0.165, trim = 20 0 0 0]{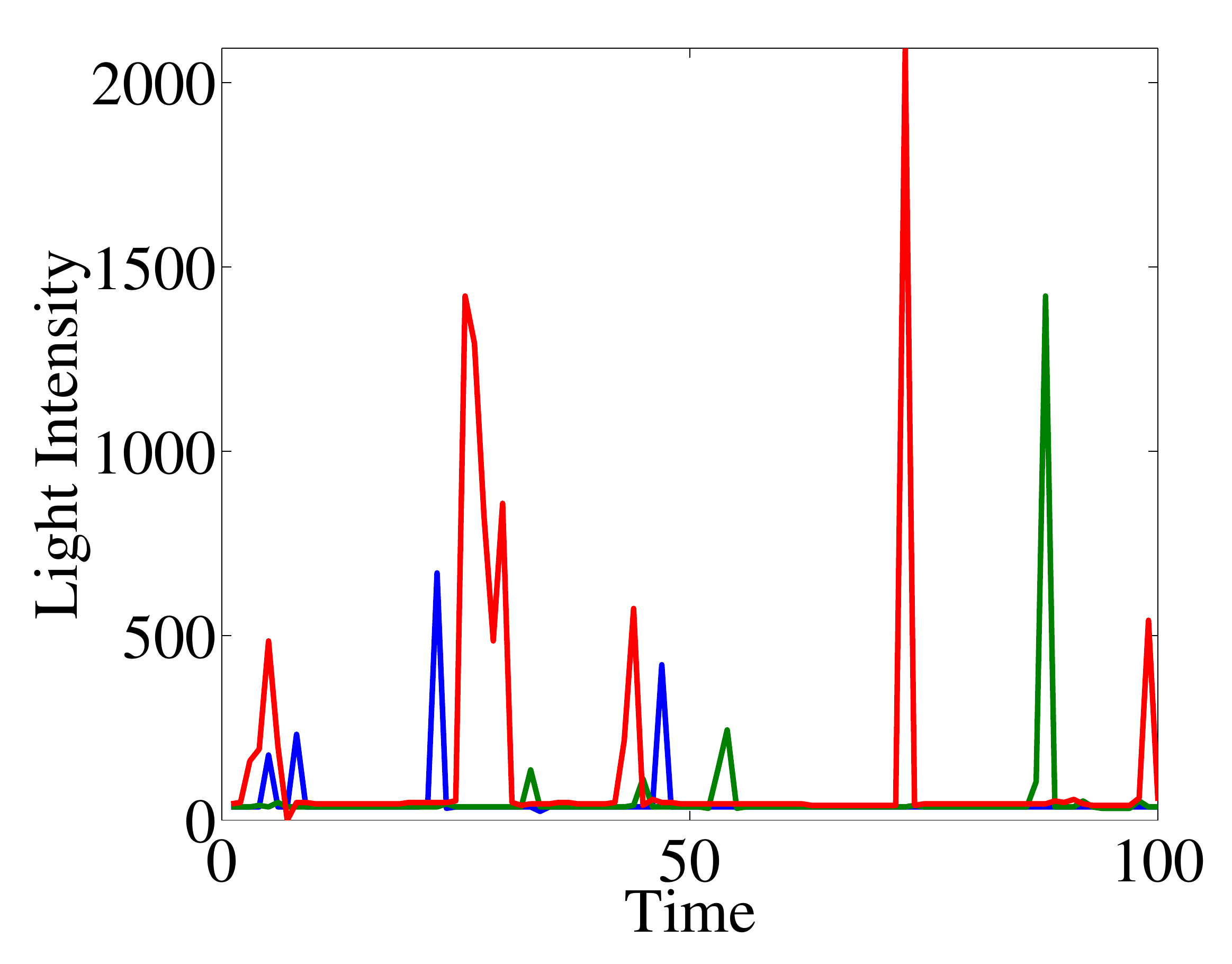}
            %  \caption{\small Light}
          %\vspace{-5pt}
              \label{subfig:light_trace}
        \end{subfigure}%
        \begin{subfigure}[h] {0.20\textwidth}
              \includegraphics[scale = 0.165, trim = 15 0 0 0]{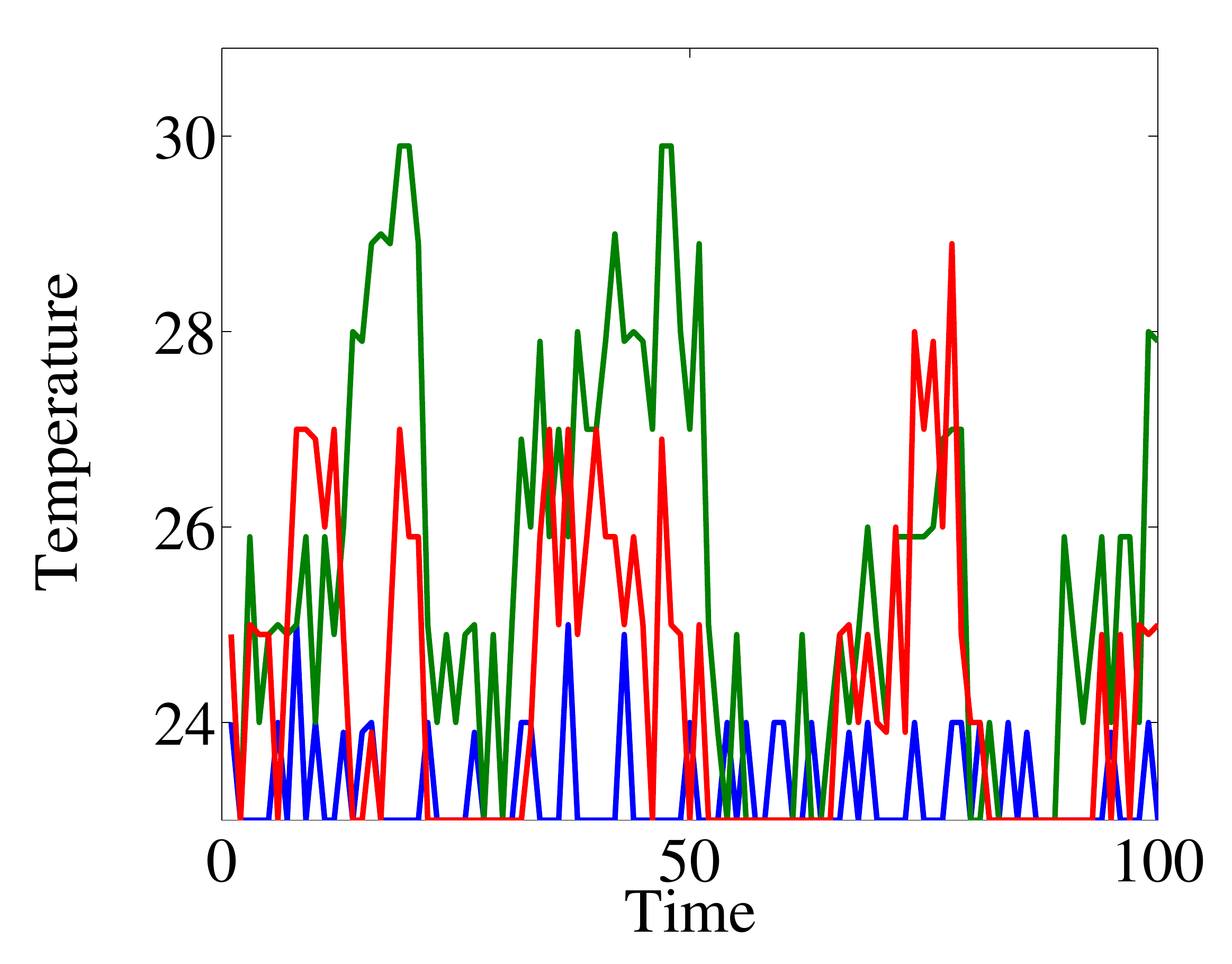}
            %  \caption{\small Temperature}
           %\vspace{-5pt}
              \label{subfig:temp_trace}
        \end{subfigure}%
        \begin{subfigure}[h] {0.20\textwidth}
              \includegraphics[scale = 0.165, trim = 10 0 0 0]{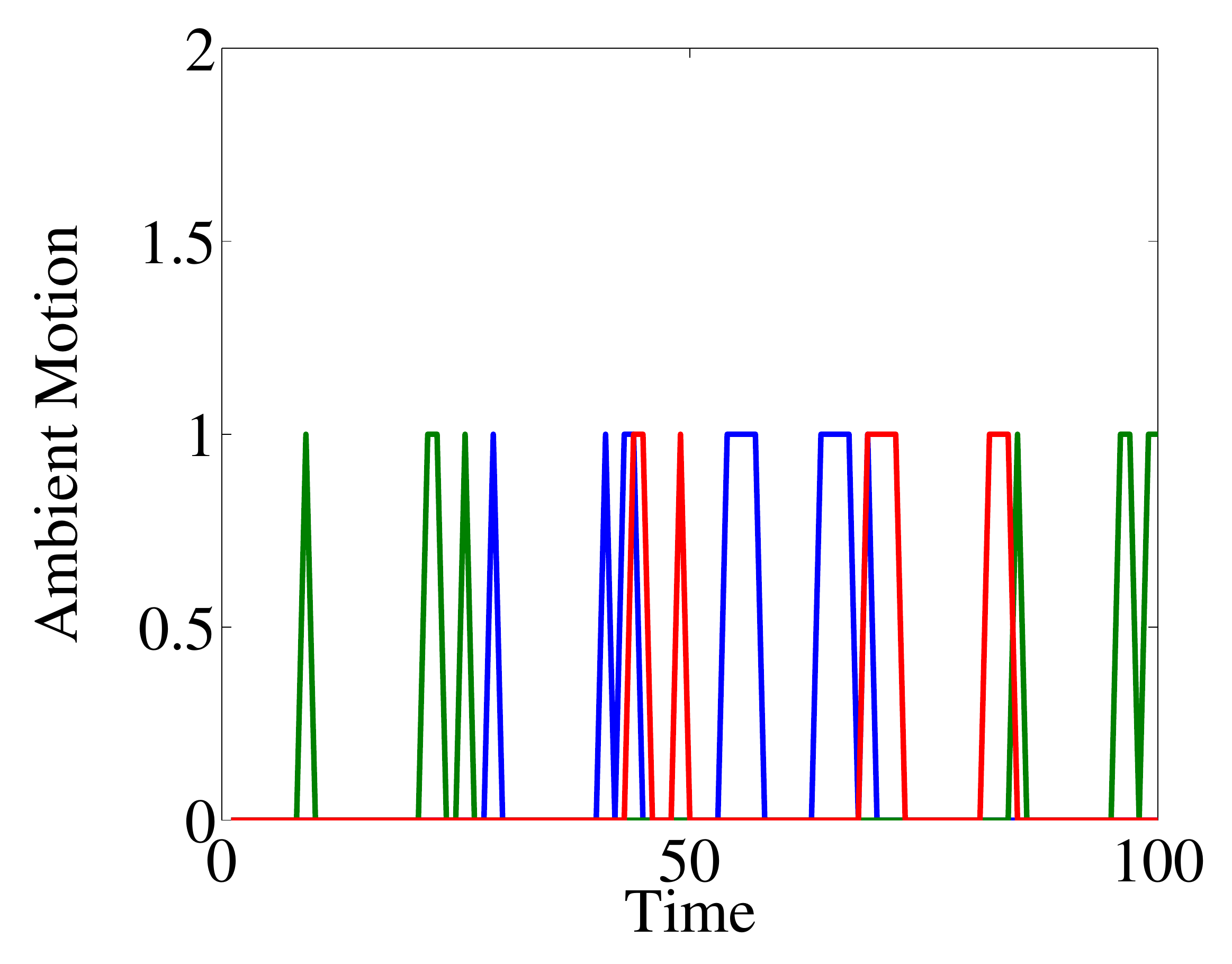}
             % \caption{\small PIR motion}
            %\vspace{-5pt}
              \label{subfig:pir_trace}
        \end{subfigure}%
        \begin{subfigure}[h] {0.20\textwidth}
              \includegraphics[scale = 0.165, trim = 0 0 0 0]{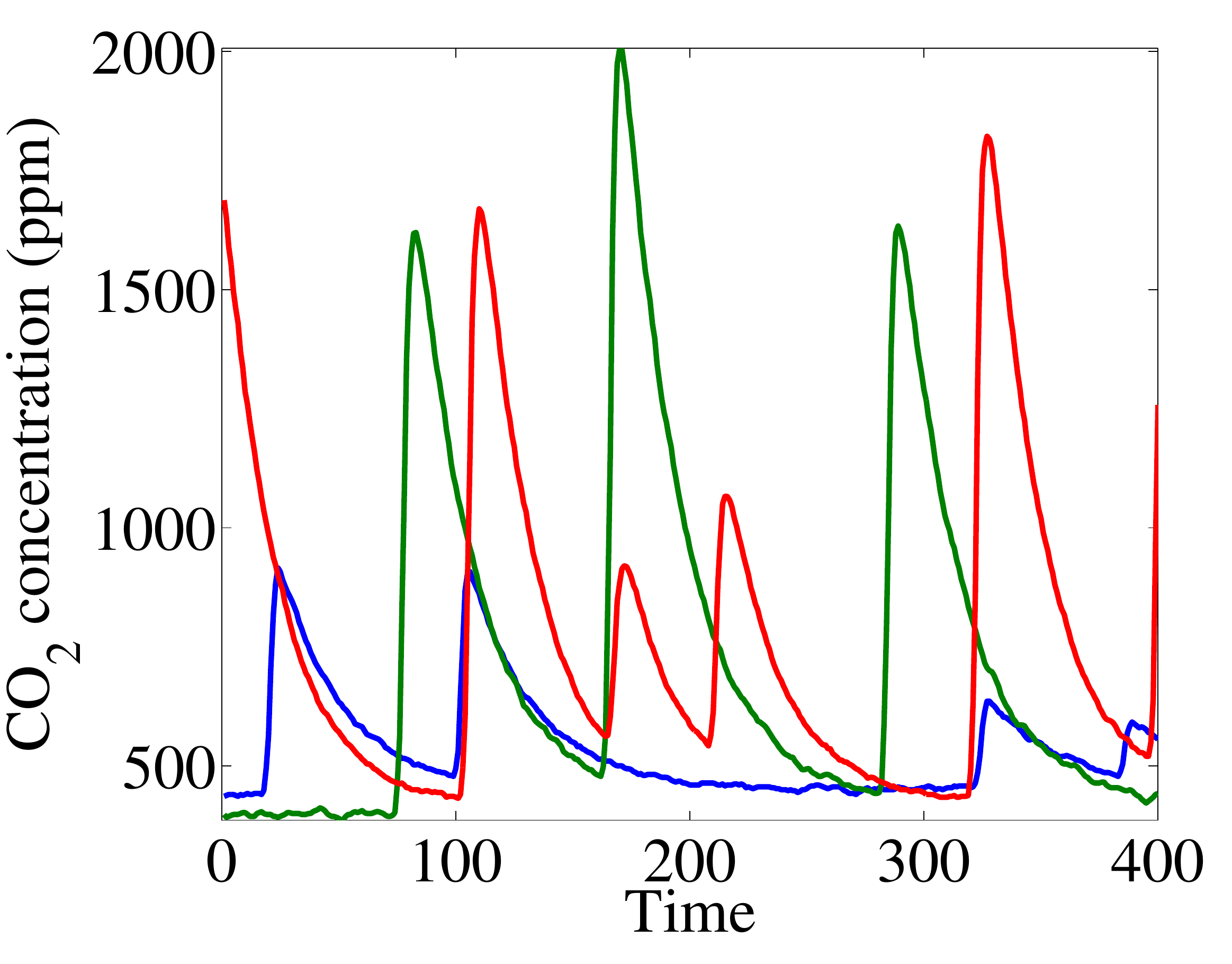}
            %  \caption{\small CO$_2$}
           %\vspace{-5pt}
              \label{subfig:co2_trace}
        \end{subfigure}%
        \begin{subfigure}[h] {0.20\textwidth}
              \includegraphics[scale = 0.165, trim = 0 0 0 0]{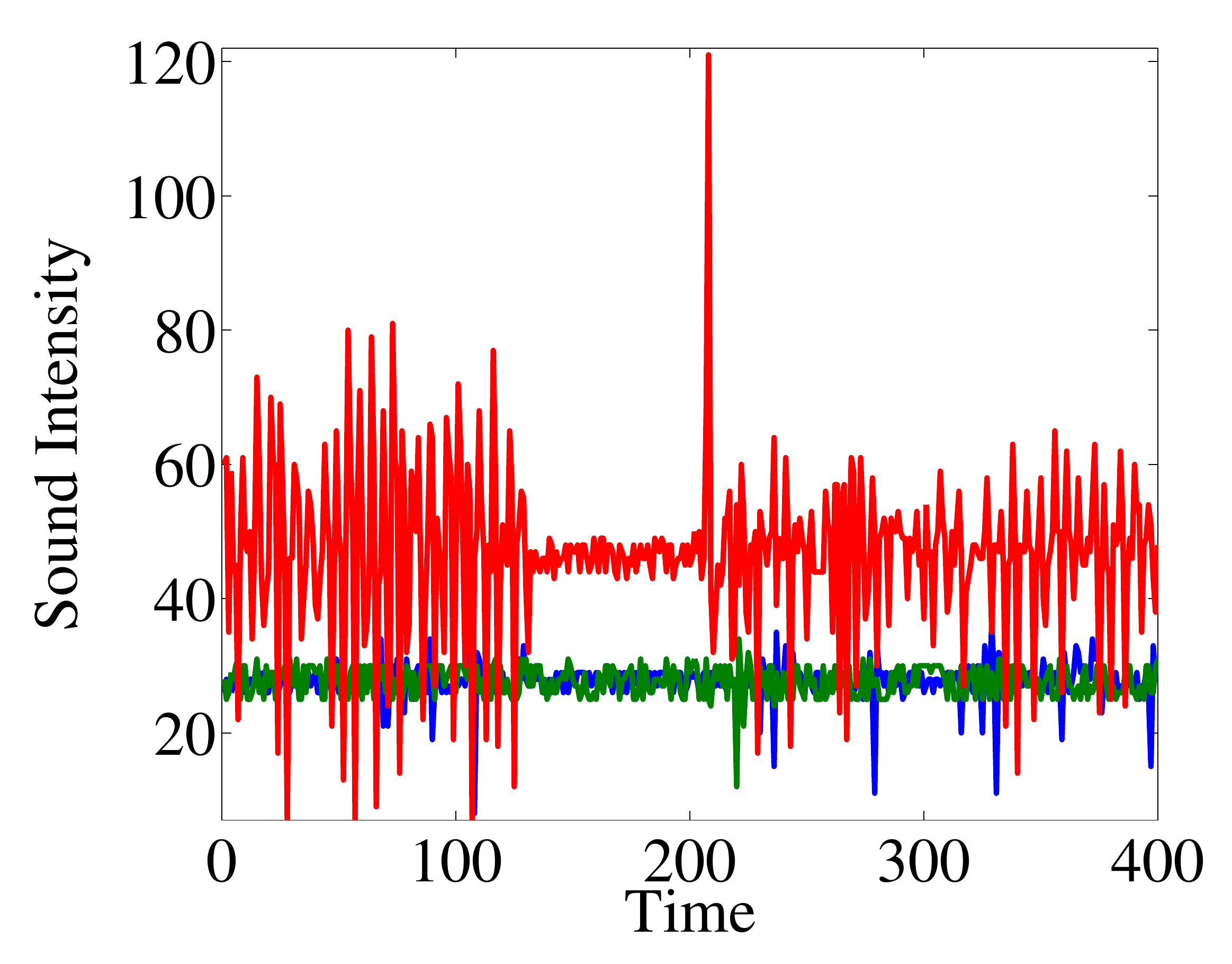}
            %  \caption{\small Sound}
           %\vspace{-5pt}
              \label{subfig:sound_trace}
        \end{subfigure}
        
     \hspace{0pt}    \begin{subfigure}[h]       
      {0.20\textwidth}
              \includegraphics[scale = 0.156, trim = 0 0 0 0]{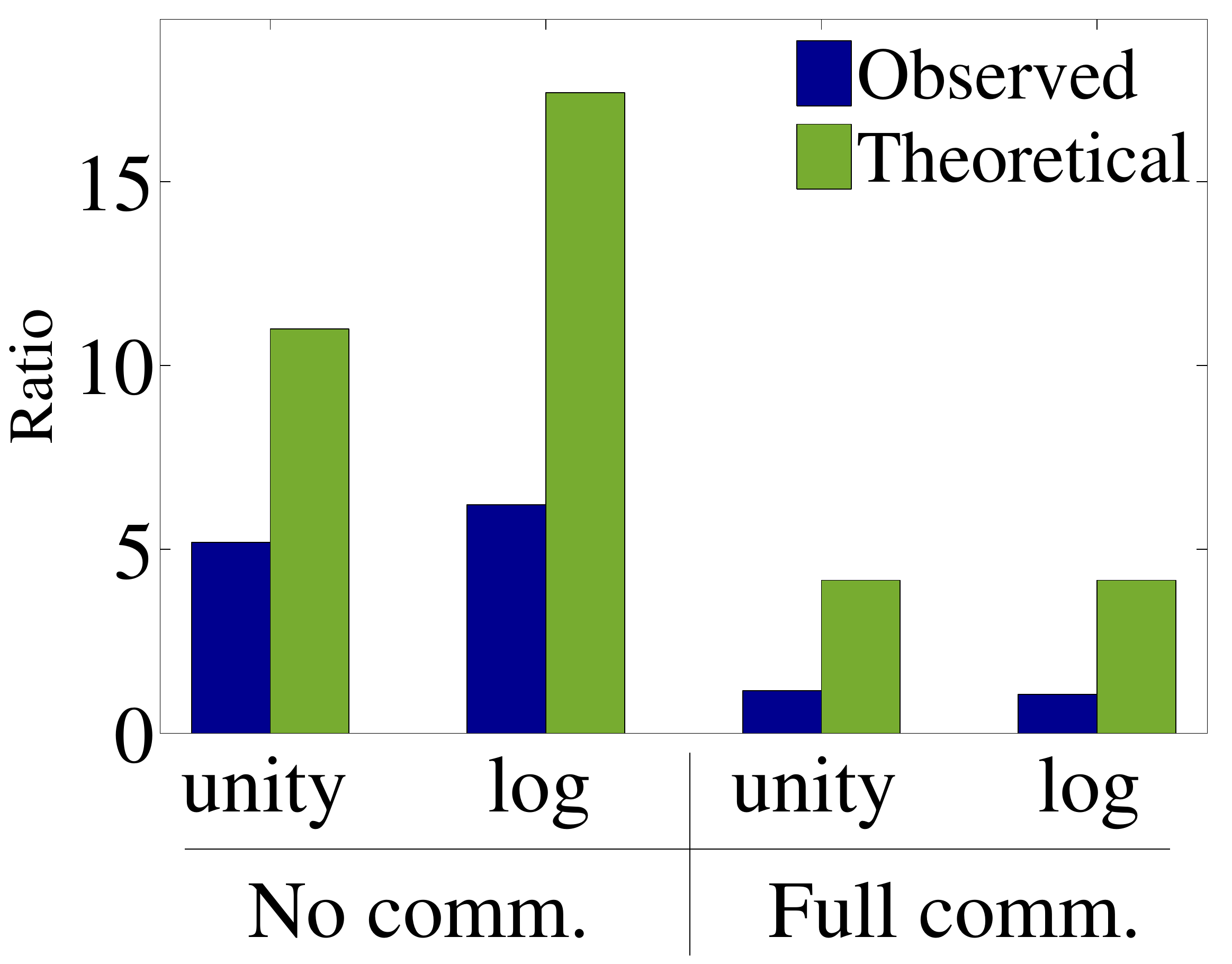}
              \caption{\small Light}
              \label{subfig:light_res}
        \end{subfigure}%
        \begin{subfigure}[h] {0.20\textwidth}
              \includegraphics[scale = 0.156, trim = 0 0 0 0]{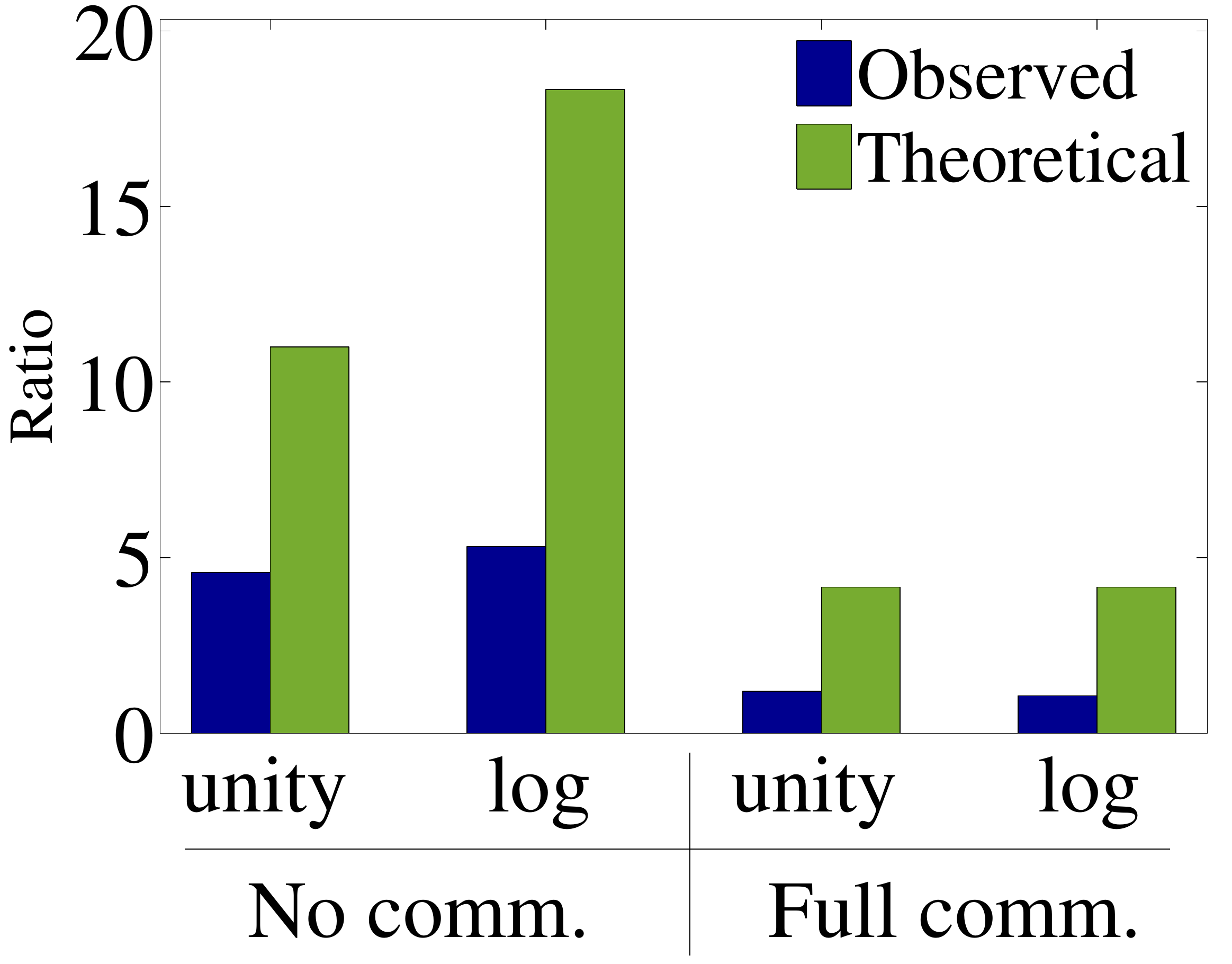}
              \caption{\small Temperature}
              \label{subfig:temp_res}
        \end{subfigure}%
        \begin{subfigure}[h] {0.20\textwidth}
              \includegraphics[scale = 0.156, trim = 0 0 0 0]{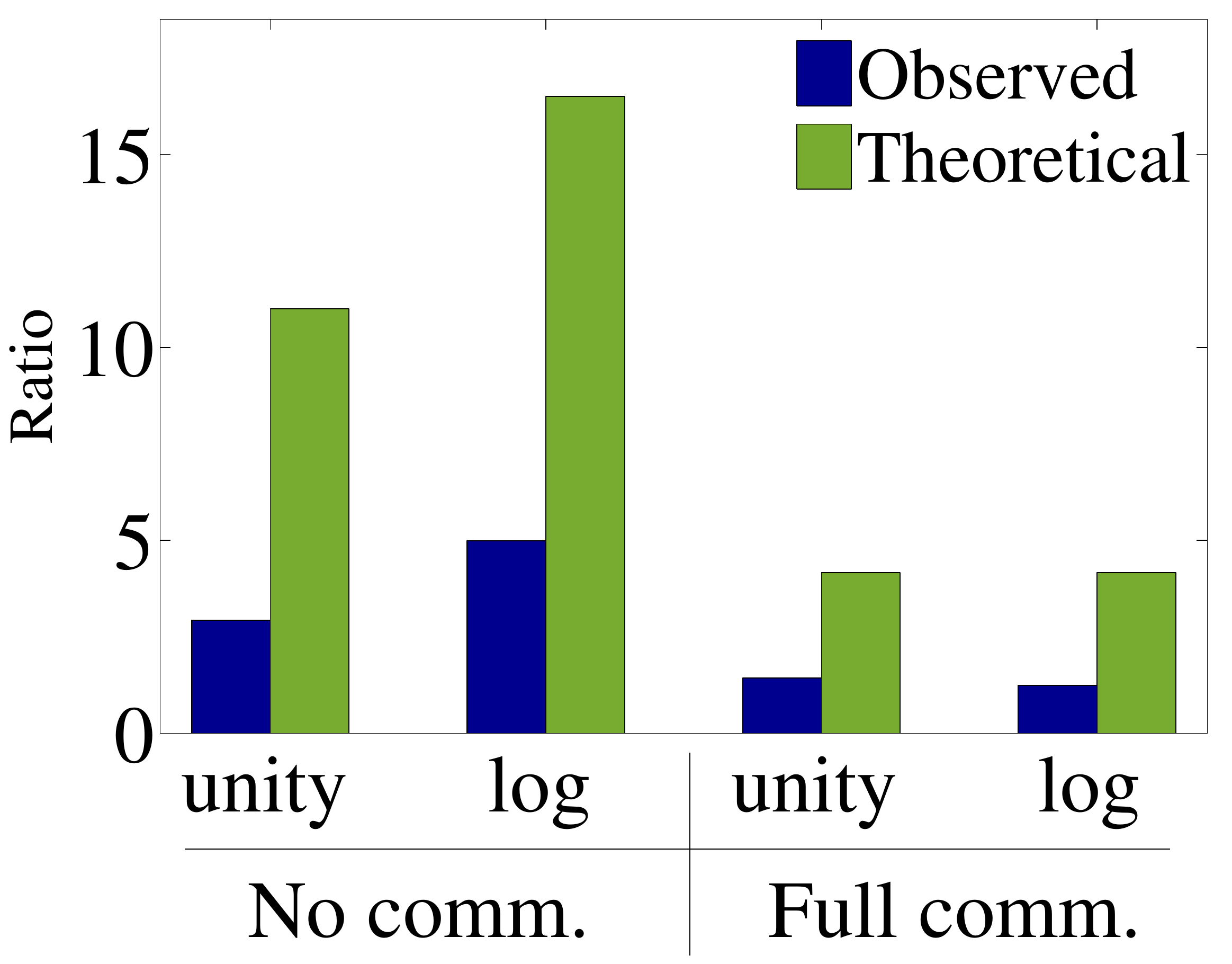}
              \caption{\small PIR motion}
              \label{subfig:pir_res}
        \end{subfigure}%
        \begin{subfigure}[h] {0.20\textwidth}
              \includegraphics[scale = 0.156, trim = 0 0 0 0]{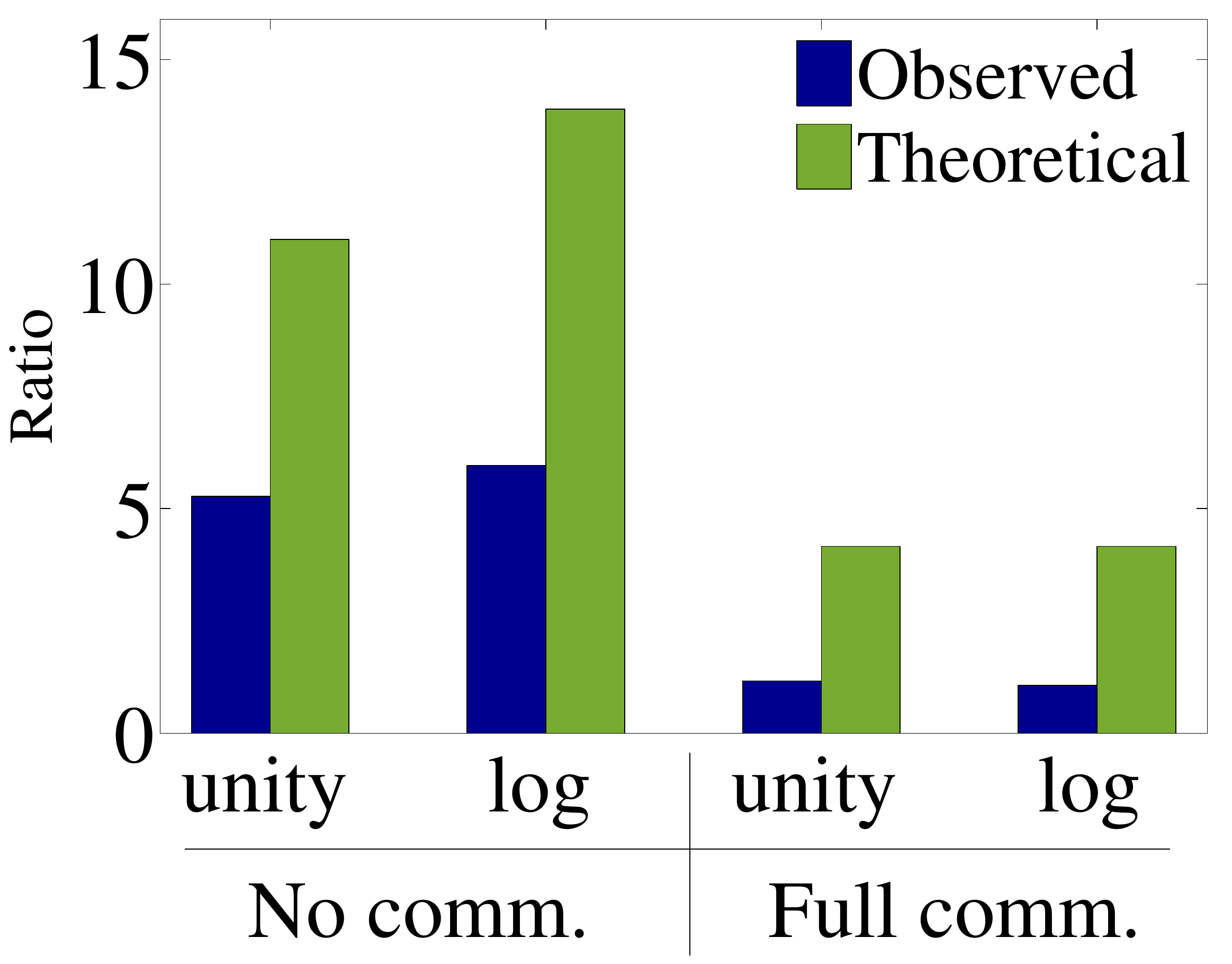}
              \caption{\small CO$_2$}
              \label{subfig:co2_res}
        \end{subfigure}%
        \begin{subfigure}[h] {0.20\textwidth}
              \includegraphics[scale = 0.156, trim = 0 0 0 0]{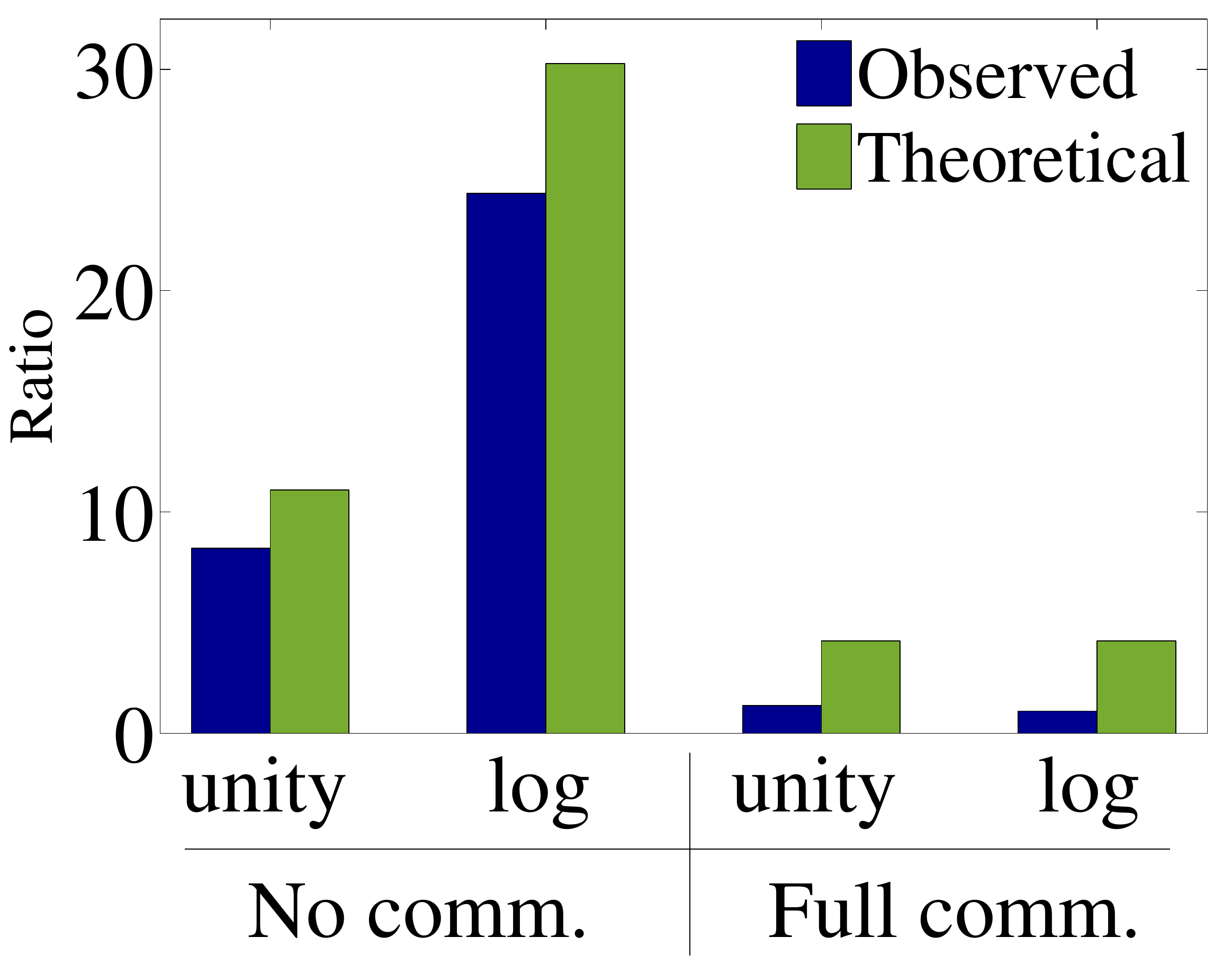}
              \caption{\small Sound}
              \label{subfig:sound_res}
        \end{subfigure}%
		\caption{\small Testbed results}\label{fig:sensor_res} \label{fig:sensor_traces}
 \end{figure*}

Sound sensors can detect the intensity of loudness of environmental sound and generate an output in the range (0-1023) accordingly. Light sensors can detect the intensity of brightness in the range (0-4015). Motion sensors can generate binary states (i.e., 1 whenever motion is detected, 0 otherwise). Figs.~\ref{subfig:light_res}, \ref{subfig:pir_res} and \ref{subfig:sound_res} show the sample traces of the sound, light, and motion sensors respectively. We observe that these sensors have rather instantaneous response time, which makes them suitable for intrusion detection applications.

Figs.~\ref{subfig:temp_res} and \ref{subfig:co2_res} show the signal traces for temperature and CO$_2$ sensors. Unlike the previous sensors, the response time of both temperature and CO$_2$ sensors is not slower due to the slowly varying concentration of CO$_2$ level. We observe that in the absence of external CO$_2$ sources (e.g., human), the values measured by the sensors are around 500 ppm, which is close to the air concentration of CO$_2$ under normal conditions. 

\smallskip

\subsubsection{Sensor Energy Consumption}

One of the design goals in wireless sensor networks is prolong the battery lifetime. The abstract communication cost in our model can naturally capture the energy consumption in communication operations. To measure the energy consumption, we conducted a sensor energy monitoring experiment, using a custom-built hardware module that monitors the energy consumption of the wireless sensors used in our experiments. The Arduino-based hardware module measures the energy consumption of various operations performed by wireless sensors such as communication, sensing, and processing. The energy consumption due to communication was measured to be 10 micro-Joules per bit, whereas the energy consumed by sensing and processing operations was observed to be negligible. In our experiments, the radio packet size was fixed at 10 bytes, which means that the wireless sensors consume about 0.8 milli-Joules energy per radio transmission. %We include the energy consumption in communication cost function.
%Table.~\ref{tab:sensor_energy} shows detailed cost information including energy cost. The data shown is for Light sensor with full intercommunication.

\smallskip

\iffalse
\begin{table}[htb!] \centering
{ 
\begin{tabularx}{\linewidth}{@{}m{4cm}| @{} m{2.35cm} @{}| @{} m{2.15cm} @{} @{} m{0cm} @{}}
%			\begin{tabularx}{\linewidth}{@{}p{1.5cm}| @{} x{.16\textwidth} @{}| @{} x{.16\textwidth} @{}| @{} x{.16\textwidth} @{}| @{} x{.16\textwidth} @{} @{} x{0.0001\textwidth} @{}}

			  \hline  \hline   
			%\multirow{2}{*}{}  & \multicolumn{2}{l|}{\textbf{No comm}} & \multicolumn{2}{l}{\textbf{Full comm}}  &\\  \cline{2-5}
			%
			 			   & \textbf{Unity Function} & \textbf{Log Function}  &\\ [0.7ex]  \hline
			 Message Cost & 11.50 & 19.21  &\\  \hline
			 Energy Cost (milli-Joule) & 9.20 & 15.37  &\\  \hline
			 Latency Cost & 16.32 & 29.17  &\\  \hline 	 Total Cost (Message+Latency) & 27.82	 & 48.38  &\\  \hline \hline
\end{tabularx} }
\caption{\small }
\label{tab:sensor_energy}
\end{table}
\fi

%%\vspace{-5pt}
\subsubsection{Empirical Results}

Figs.~\ref{subfig:light_res}-\ref{subfig:sound_res} show the observed ratios achieved by our threshold-based online algorithms in the sensor network testbed. There are several observations:
\begin{itemize}
\item The observed ratio in each case is considerably below the corresponding theoretical competitive ratio. 
\item The observed ratio achieved in the setting of full intercommunication is better than the setting of no intercommunication, because in the former case the number of transmissions by the online algorithm is near optimal as all sensors can overhear each other radio transmissions. In the latter case, on the other hand, the algorithm tends to transmit more reports because the sensors cannot overhear each other. 
\item Unity cost results in a lower observed ratio than logarithmic cost which is verified by our simulation results.
\end{itemize}

\smallskip

%\vspace{-5pt}
\section{Conclusion} \label{sec:disc}

In this paper, we shed light on the distributed online optimization of cost efficiency and effectiveness of information aggregation. In particular, we present competitive online algorithms and theoretical analysis for the trade-off optimization of general communication cost and latency penalty with distributed systems in various settings of intercommunication. All the proofs can be found in the technical report \cite{CKAToN:2016}.

In a companion paper \cite{Khonji:2014}, we also study randomized online algorithms to solve the distributed online optimization, which relies on a different formulation using LP approach. 

Distributed online algorithms are a burgeoning area, which is vital to many applications in networking and distributed systems. Our study concerns a general problem, paving the way of tackling a wide range of distributed online problems.

\bibliographystyle{IEEEtran}
\bibliography{reference}

\iffalse
\vspace{-50pt}
\begin{IEEEbiographynophoto}
{Chi-Kin Chau} is a faculty member with the Department of EECS at Masdar Institute of Science and Technology.
\end{IEEEbiographynophoto}

\vspace{-50pt}
\begin{IEEEbiographynophoto}
{Majid Khonji} is a Ph.D. student with the Department of EECS at Masdar Institute of Science and Technology.
\end{IEEEbiographynophoto}

\vspace{-50pt}
\begin{IEEEbiographynophoto}
{Muhammad Aftab} is a Ph.D. student with the Department of EECS at Masdar Institute of Science and Technology.
\end{IEEEbiographynophoto}
\fi

\clearpage

%\vspace{-5pt}
\appendix

%\vspace{-5pt}

\subsection{Proofs}

\begin{lemma} \label{lem:subadd}
(Subadditivity) For a set of segments ${\cal S}$,
\begin{equation} \label{eqn:subadd}
 {\tt Cost}({\cal A}_{\rm thb}[\theta, \sigma]) \le \sum_{{\tt S} \in {\cal S}} {\tt Cost}({\cal A}_{\rm thb}[\theta, \sigma_{\tt S}]) 
\end{equation}
Subadditivity property (Eqn.~(\ref{eqn:subadd})) also holds for threshold-based algorithms ${\cal A}^{{\rm itc.}K}_{\rm thb}$ and ${\cal A}^{{\rm net.}K}_{\rm thb}$.
\end{lemma}

\begin{IEEEproof}
It suffices to consider one system. Without loss of generality, we consider a pair of consecutive segments ${\tt S}_1, {\tt S}_2 \in {\cal S}$. Let ${\tt Cost}({\cal A}_{\rm thb}[\theta, \sigma_{{\tt S}_1 \cup {\tt S}_2}])= \rho {\cal C}_{{\tt S}_1 \cup {\tt S}_2} + (1-\rho) {\cal L}_{{\tt S}_1 \cup {\tt S}_2}$, ${\tt Cost}({\cal A}_{\rm thb}[\theta, \sigma_{{\tt S}_1}])= \rho {\cal C}_{{\tt S}_1} + (1-\rho) {\cal L}_{{\tt S}_1}$ and ${\tt Cost}({\cal A}_{\rm thb}[\theta, \sigma_{{\tt S}_2}])= \rho {\cal C}_{{\tt S}_2} + (1-\rho) {\cal L}_{{\tt S}_2}$.

Note that the communication cost is sub-additive. Thus,
\begin{equation}
{\cal C}_{{\tt S}_1 \cup {\tt S}_2} \le {\cal C}_{{\tt S}_1} + {\cal C}_{{\tt S}_2}
\end{equation}

By threshold-based algorithm, one obtains 
\begin{equation}
\frac{{\cal L}_{{\tt S}_1 \cup {\tt S}_2}}{{\cal C}_{{\tt S}_1 \cup {\tt S}_2}} = \frac{{\cal L}_{{\tt S}_1}}{{\cal C}_{{\tt S}_1}} = \frac{{\cal L}_{{\tt S}_2}}{{\cal C}_{{\tt S}_2}} = \theta
\end{equation}
Hence,
\begin{equation}
{\cal L}_{{\tt S}_1 \cup {\tt S}_2} = \theta {\cal C}_{{\tt S}_1 \cup {\tt S}_2} \le
\theta ({\cal C}_{{\tt S}_1} + {\cal C}_{{\tt S}_2}) = {\cal L}_{{\tt S}_1} + {\cal L}_{{\tt S}_2}
\end{equation}
Therefore, 
\begin{equation*}
{\tt Cost}({\cal A}_{\rm thb}[\theta, \sigma_{{\tt S}_1 \cup {\tt S}_2}]) \le {\tt Cost}({\cal A}_{\rm thb}(\theta, \sigma_{{\tt S}_1})) + {\tt Cost}({\cal A}_{\rm thb}[\theta, \sigma_{{\tt S}_2}])
\end{equation*}
Similarly, the preceding argument can be applied to threshold-based algorithms ${\cal A}^{{\rm itc.}K}_{\rm thb}$ and ${\cal A}^{{\rm net.}K}_{\rm thb}$.
\end{IEEEproof}

\smallskip

\begin{customthm}{2} 
For {\rm $K$-DIA}, let threshold $\theta = {\textstyle\frac{K \rho}{\alpha N(1-\rho)}}$ for every system $i \in {\cal N}$, the competitive ratio of ${\cal A}_{\rm thb}$ is ${\tt CR}({\cal A}_{\rm thb}[{\textstyle\frac{K \rho}{\alpha N(1-\rho)}}]) = \frac{\alpha N}{K}+1$.
\end{customthm}
\begin{IEEEproof}
Let ${\tt Opt}_K(\sigma)$ be the $K$-report optimal offline solution, and its total cost be $\rho {\cal C}^{\ast}_K + (1-\rho) {\cal L}^{\ast}_K$. 

Next, consider a different communication cost function, defined by $\tilde{\cal C}({\sigma}_{i,k}) \triangleq K\cdot{\cal C}({\sigma}_{i,k})$. Let $\widetilde{\tt Opt}(\sigma)$ be the $1$-report optimal offline solution using communication cost function $\tilde{\cal C}({\sigma}_{i,k})$, and its total cost be $\rho \tilde{\cal C}^{\ast} + (1-\rho) \tilde{\cal L}^{\ast}$. 

$\widetilde{\tt Opt}(\sigma)$ can be compared with ${\tt Opt}_1(\sigma)$, such that each report message is replicated $K$ times from the same system that has the minimal communication cost. On the other hand, ${\tt Opt}_K(\sigma)$ requires $K$ report messages from different systems.
Since $K$ messages can be transmitted at the system with the least communication cost for $\widetilde{\tt Opt}(\sigma)$, we obtain
\begin{equation}
\rho {\cal C}^{\ast}_K + (1-\rho) {\cal L}^{\ast}_K \ge
\rho \tilde{\cal C}^{\ast} + (1-\rho) \tilde{\cal L}^{\ast} 
\end{equation}

Therefore, the competitive ratio of ${\cal A}_{\rm thb}$ is bounded by	
\begin{equation}
\frac{ {\tt Cost}({\cal A}_{\rm thb}[{\textstyle\frac{K\rho}{\alpha N(1-\rho)}}, \sigma]) }{{\tt Opt}_K(\sigma)} \le
\frac{ {\tt Cost}({\cal A}_{\rm thb}[{\textstyle\frac{K\rho}{\alpha N(1-\rho)}}, \sigma]) }{\widetilde{\tt Opt}(\sigma)} 
\end{equation}

In the following, we consider $\widetilde{\tt Opt}(\sigma)$. Note that ${\cal A}_{\rm thb}$ behaves similarly for {\rm $K$-DIA}.
As in Theorem~\ref{thm:woc.alg}, we define ${\cal S}^\ast$, ${\cal G}^{(1)}$, ${\cal G}^{(2)}$, ${\cal C}^{(1)}$, ${\cal L}^{(1)}$, ${\cal C}^{(2)}$, and ${\cal L}^{(2)}$. Next, the competitive ratio is obtained with respect to each $\sigma_{\tt S}$ for ${\tt S} \in {\cal S}^\ast$. By a slight abuse of notation, let the total cost of $\widetilde{\tt Opt}(\sigma_{\tt S})$ be $\rho \tilde{\cal C}^{\ast} + (1-\rho) \tilde{\cal L}^{\ast}$.

For ${\cal G}^{(1)}$, as in Theorem~\ref{thm:woc.alg}, we obtain
\begin{equation}
{\tt Cost}({\cal A}_{\rm thb}[{\textstyle\frac{K\rho}{\alpha N(1-\rho)}}, {\cal G}^{(1)}]) \le (1-\rho)(\frac{\alpha N}{K}+1) \tilde{\cal L}^\ast
\end{equation}
However, for ${\cal G}^{(2)}$, we obtain $\frac{\alpha N}{K}  \tilde{\cal C}^{\ast} \ge  {\cal C}^{(2)}$.
Therefore,
\begin{equation}
{\tt Cost}({\cal A}_{\rm thb}[{\textstyle\frac{K\rho}{\alpha N(1-\rho)}}, {\cal G}^{(2)}]) \le \rho(\frac{\alpha N}{K}+1) \tilde{\cal C}^\ast
\end{equation}
Finally, it follows that
\begin{eqnarray}
&& \frac{ {\tt Cost}({\cal A}_{\rm thb}[{\textstyle\frac{K\rho}{\alpha N(1-\rho)}}, \sigma_{\tt S}]) }{\widetilde{\tt Opt}(\sigma_{\tt S})} \\ 
& \le & \frac{(1-\rho) (\frac{\alpha N}{K}+1) \tilde{\cal L}^{\ast} + \rho (\frac{\alpha N}{K}+1) \tilde{\cal C}^{\ast}}{\rho \tilde{\cal C}^{\ast} + (1-\rho) \tilde{\cal L}^{\ast}}  =  \frac{\alpha N}{K}+1
\end{eqnarray}
\end{IEEEproof}

\smallskip

\begin{customthm}{4} 
For {\rm $K$-DIA}, let threshold $\theta = {\textstyle\frac{\rho}{(1-\rho)\phi}}$ for every system $i \in {\cal N}$ where $\phi = \frac{\sqrt{(\alpha-K)^2 + 4 \alpha K N} + \alpha - K}{2 K}$, the competitive ratio ${\tt CR}({\cal A}^{{\rm itc.}K}_{\rm thb}[{\textstyle\frac{\rho}{(1-\rho)\phi}}]) = \frac{\sqrt{(\alpha-K)^2 + 4 \alpha K N} + \alpha + K}{2 K}$.
\end{customthm}
\begin{IEEEproof}
As in Theorem~\ref{thm:woc.alg.k}, we define $\tilde{\cal C}({\sigma}_{i,k}) \triangleq K\cdot{\cal C}({\sigma}_{i,k})$. Let $\widetilde{\tt Opt}(\sigma)$ be the $1$-report optimal offline solution using communication cost function $\tilde{\cal C}({\sigma}_{i,k})$. 
As in Theorem~\ref{thm:woc.alg}, we also define ${\cal S}^\ast$ with respect to $\widetilde{\tt Opt}(\sigma)$, $\tilde{\cal C}^{\ast}$, $\tilde{\cal L}^{\ast}$, ${\cal G}^{(1)}$, ${\cal G}^{(2)}$, ${\cal C}^{(1)}$, ${\cal L}^{(1)}$ and ${\cal L}^{(2)}$. 
Next, the competitive ratio is obtained with respect to each $\sigma_{\tt S}$ for ${\tt S} \in {\cal S}^\ast$. 

As in Theorem~\ref{thm:whc.alg}, the total cost of ${\cal A}^{{\rm itc.}K}_{\rm thb}$ for ${\cal G}^{(1)}$ is 
\begin{equation}
{\tt Cost}({\cal A}^{{\rm itc.}K}_{\rm thb}[{\textstyle\frac{\rho}{(1-\rho)\phi}}, {\cal G}^{(1)}]) \le (1-\rho)(\phi + 1) \tilde{\cal L}^\ast
\end{equation}

Let the number of report messages for ${\cal G}^{(2)}$ be $\eta^{(2)}$ and the $k$-th report communication cost be ${\cal C}^{(2)}_k$. Since $\widetilde{\tt Opt}(\sigma)$ generates only one report message at communication cost $\tilde{\cal C}^{\ast}$, which is $K$-time of the per message communication cost of ${\cal A}^{{\rm itc.}K}_{\rm thb}$, it follows that $\max_{k \in \{1,...,\eta^{(2)}\}} {\cal C}^{(2)}_k \le \frac{\alpha\tilde{\cal C}^{\ast}}{K}$, and
\begin{equation}
{\frac{(1-\rho)\phi}{\rho}} {\cal L}^{(2)} = \sum_{k=1}^{\eta^{(2)}} {\cal C}^{(2)}_k \le \alpha N {\frac{\tilde{\cal C}^{\ast}}{K}}
\end{equation}
Therefore, the total cost of ${\cal A}^{{\rm itc.}K}_{\rm thb}$ for ${\cal G}^{(2)}$ is bounded by
\begin{equation}
{\tt Cost}({\cal A}^{{\rm itc.}K}_{\rm thb}[{\textstyle\frac{\rho}{(1-\rho)\phi}}, {\cal G}^{(2)}]) \le \frac{\alpha \rho}{K} ({\frac{N}{\phi}} + 1) \tilde{\cal C}^{\ast}
\end{equation}

To sum up, the competitive ratio is obtained as follows.
\begin{eqnarray}
\displaystyle
& & \frac{ {\tt Cost}({\cal A}^{{\rm itc.}K}_{\rm thb}[{\textstyle\frac{\rho}{(1-\rho)\phi}}, \sigma_{\tt S}]) }{{\tt Opt}(\sigma_{\tt S})} \\
 & \le &
\displaystyle \frac{(1-\rho) (\phi + 1) {\cal L}^{\ast} + \frac{\alpha \rho}{K} ({\frac{N}{\phi}} + 1) \tilde{\cal C}^{\ast}}{\rho \tilde{\cal C}^{\ast} + (1-\rho) \tilde{\cal L}^{\ast}} \\
& = & \phi + 1 = \frac{\sqrt{(\alpha-K)^2 + 4 \alpha K N} + \alpha + K}{2 K}
\end{eqnarray}
The last equality is because that $\phi = \frac{\sqrt{(\alpha-K)^2 + 4 \alpha K N} + \alpha - K}{2 K}$ is the positive root of $\phi$ in $\phi + 1 = \frac{\alpha \rho}{K} ({\frac{N}{\phi}} + 1)$.
\end{IEEEproof}

\smallskip

\begin{customthm}{5} 
Given a communication graph ${\cal G}$, with ${\cal N}_{\rm w}$ and ${\cal N}_{\rm f}$, Setting threshold $\theta = {\textstyle\frac{\rho}{(1-\rho)\phi}}$ for every system $i \in {\cal N}$, where 
\begin{equation}
\phi = \frac{\sqrt{(\alpha {\rm x} -K)^2 + 4 \alpha K N} + \alpha {\rm x} - K}{2 K}
\end{equation}
the competitive ratio of ${\cal A}^{{\rm net.}K}_{\rm thb}$ is 
\begin{equation}
{\tt CR}({\cal A}^{{\rm net.}K}_{\rm thb}[{\textstyle\frac{\rho}{(1-\rho)\phi}}]) = \frac{\sqrt{(\alpha {\rm x} - K)^2 + 4 \alpha K N} + \alpha {\rm x} + K}{2 K} 
\end{equation}
\end{customthm}

\begin{IEEEproof}
The proof extends the arguments of Theorem~\ref{thm:whc.alg.k}. 

As in Theorem~\ref{thm:whc.alg}, we define $\tilde{\cal C}({\sigma}_{i,k}) \triangleq K\cdot{\cal C}({\sigma}_{i,k})$. Let $\widetilde{\tt Opt}(\sigma)$ be the $1$-report optimal offline solution using communication cost function $\tilde{\cal C}({\sigma}_{i,k})$. 
As in Theorem~\ref{thm:woc.alg}, we define ${\cal S}^\ast$ with respect to $\widetilde{\tt Opt}(\sigma)$, $\tilde{\cal C}^{\ast}$, $\tilde{\cal L}^{\ast}$, ${\cal G}^{(1)}$, ${\cal G}^{(2)}$, ${\cal C}^{(1)}$, ${\cal L}^{(1)}$ and ${\cal L}^{(2)}$. Next, the competitive ratio is obtained with respect to each $\sigma_{\tt S}$ for ${\tt S} \in {\cal S}^\ast$.

First, we obtain
\begin{equation}
{\cal C}^{(1)}  = {\frac{(1-\rho)\phi}{\rho}} {\cal L}^{(1)} \le {\frac{(1-\rho)\phi}{\rho}} \tilde{\cal L}^\ast
\end{equation}
Therefore, the total cost of ${\cal A}^{{\rm net.}K}_{\rm thb}$ for ${\cal G}^{(1)}$ is bounded by
\begin{equation}
{\tt Cost}({\cal A}^{{\rm net.}K}_{\rm thb}[{\textstyle\frac{\rho}{(1-\rho)\phi}}, {\cal G}^{(1)}]) \le (1-\rho)(\phi + 1) \tilde{\cal L}^\ast
\end{equation}

Let the number of report messages for ${\cal G}^{(2)}$ be $\eta^{(2)}$ and the $k$-th report communication cost be ${\cal C}^{(2)}_k$. As in Theorem~\ref{thm:whc.alg.k}, it follows that $\max_{k \in \{1,...,\eta^{(2)}\}} {\cal C}^{(2)}_k \le \frac{\alpha\tilde{\cal C}^{\ast}}{K}$, and
\begin{equation}
{\frac{(1-\rho)\phi}{\rho}} {\cal L}^{(2)} \le \sum_{k=1}^{\eta^{(2)}} {\cal C}^{(2)}_k \le \alpha N {\frac{\tilde{\cal C}^{\ast}}{K}}
\end{equation}
On the other hand, since the maximum number of systems that generate simultaneous report messages for ${\cal G}^{(2)}$ is ${\rm x}$, it follows that $\alpha {\rm x} \tilde{\cal C}^{\ast}  \ge \sum_{k = 1}^{\eta^{(2)}} {\cal C}^{(2)}_k$.

Therefore, the total cost of ${\cal A}^{{\rm net.}K}_{\rm thb}$ for ${\cal G}^{(2)}$ is bounded by
\begin{equation}
{\tt Cost}({\cal A}^{{\rm net.}K}_{\rm thb}[{\textstyle\frac{\rho}{(1-\rho)\phi}}, {\cal G}^{(2)}]) \le \alpha \rho(\frac{N}{K \phi} + {\rm x})\tilde{\cal C}^{\ast}
\end{equation}

Finally, it follows that
\begin{eqnarray}
&&\frac{ {\tt Cost}({\cal A}^{{\rm net.}K}_{\rm thb}[\frac{\rho \phi}{1-\rho}, \sigma_{\tt S}]) }{\widetilde{\tt Opt}(\sigma_{\tt S})} \\ 
& \le & \frac{(1-\rho) (\phi + 1) \tilde{\cal L}^{\ast} + \alpha \rho(\frac{N}{K \phi} + {\rm x})\tilde{\cal C}^{\ast}}{\rho \tilde{\cal C}^{\ast} + (1-\rho) \tilde{\cal L}^{\ast}} \\
& = & \phi + 1 = \frac{\sqrt{(\alpha {\rm x} - K)^2 + 4 \alpha K N} + \alpha {\rm x} + K}{2 K} 
\end{eqnarray}
The last equality is because that $\phi = \frac{\sqrt{(\alpha {\rm x} -K)^2 + 4 \alpha K N} + \alpha {\rm x} - K}{2 K}$ is the positive root of $\phi$ in $\phi + 1 = \alpha(\frac{N}{K \phi} + {\rm x})$.
\end{IEEEproof}

\smallskip

%\vspace{-5pt}

%\subsection{Lower Bounds of Competitive Ratios}
%\vspace{-5pt}

\begin{customthm}{6}
Consider {\rm $1$-DIA}. Suppose $\rho=\frac{1}{2}$ and there is one system having a communication cost per report as 1, while other $(N-1)$ systems having a communication cost per report as $\alpha \ge 1$. There exists no deterministic distributed online algorithm with no intercommunication to achieve a competitive ratio lower than $\alpha(N-1)+\frac{3}{2}$.
\end{customthm}

\begin{IEEEproof}
It suffices to consider {\rm $1$-sDIA}. 
First, we consider a threshold-based algorithm ${\cal A}$, such that each system $i$ uses threshold $\theta_i$. It will be generalized to general algorithms.

Let the system having a communication cost per report as 1 be $i^\circ$.
Consider an input $\sigma$ with two events $\{j_1, j_2\}$ of the following setting:
\begin{equation}
{\tt t}^{j_1} = 0, \quad {\tt t}^{j_2} = 1, \quad
w_i^{j_1} = w_i^{j_2} = 
\left\{
\begin{array}{rl}
\theta_i +\epsilon, & \mbox{if\ } i = i^\circ \\
\alpha\theta_i +\epsilon, & \mbox{if\ } i \ne i^\circ 
\end{array}
\right. \notag
\end{equation}
for an arbitrarily small $\epsilon$ (e.g., $\epsilon = O(\frac{1}{N})$).
With no intercommunication, ${\cal A}$ needs to produce two report messages from each system, incurring a latency penalty of at least $\theta_i$ for the first event for $i^\circ$ and $\alpha \theta_i$ for other $(N-1)$ systems. Note that $\alpha \theta_i \ge  \theta_i$. Hence, it follows that
\begin{equation}
{\tt Cost}({\cal A}[\sigma]) \ge \frac{1}{2} \Big( 2\alpha(N-1) + 2 + \sum_{i \in {\cal N}} \theta_i\Big)
\end{equation}
Recall that ${\tt Opt}(\sigma)$ is the cost of a (centralized) optimal offline solution, which knows all the inputs in advance without the need for intercommunication. 
Note that ${\tt Opt}(\sigma) = \frac{1}{2} \min\{2, 1 + \sum_{i \in {\cal N}} \theta_i\}$, which either produces two report messages, or just one report message immediately following the second event plus the total latency penalty $\sum_{i \in {\cal N}} \theta_i$ for the first event. Hence, the competitive ratio becomes
\begin{equation}
{\tt CR}({\cal A}) \ge \frac{2\alpha(N-1) + 2 + \sum_{i \in {\cal N}} \theta_i}{\min\{2, 1 + \sum_{i \in {\cal N}} \theta_i\}}
\end{equation}
Note that the denominator is $1 \le \min\{2, 1 + \sum_{i \in {\cal N}} \theta_i\} \le 2$. ${\tt CR}({\cal A})$ is minimized, when $\sum_{i \in {\cal N}} \theta_i = 1$ because
\begin{equation}
x \ge y \Rightarrow \frac{z + x}{1 + x} \le \frac{z + y}{1 + y}, \mbox{\ for all\ } x \le 1, z \ge 1
\end{equation}
Hence, it follows that ${\tt CR}({\cal A})\ge \alpha(N-1)+\frac{3}{2}$.

Next, we consider non-threshold-based algorithm ${\cal A}'$. In this case, ${\cal A}'$ may have time-varying behavior. But we assume that the behavior of ${\cal A}'$ is always identical after resetting $t=0$, because ${\cal A}'$ is deterministic. Note that the optimal offline cost ${\tt Opt}(\sigma)$ does not change.  We treat ${\cal A}'$ as a blackbox. For each system $i$, we input a measurement value $w_i=1$ at time $t=0$. Suppose that the system waits for $\tau_i$ to produce a report message. The competitive ratio of ${\cal A}'$ can be lower bounded by the one of a threshold-based algorithm ${\cal A}$ with threshold $\theta_i= \tau_i$ using the preceding argument.
\end{IEEEproof}

\smallskip

\begin{customthm}{7} 
For {\rm $1$-DIA} and $\rho=\frac{1}{2}$, there exists no deterministic distributed online algorithm with collision-free intercommunication to achieve a competitive ratio lower than $\frac{\sqrt{N}}{4}$.
\end{customthm}

\begin{IEEEproof}
{\bf Step 1}:
It suffices to consider {\rm $1$-sDIA}.
First, we consider a threshold-based algorithm ${\cal A}$, such that each system $i$ uses threshold $\theta_i$.
Let $\theta_{\min} = \min_{i \in {\cal N}} \{\theta_i\}$. We then generalize the proof to general algorithms. Note that it suffices to consider the case when $\frac{\sqrt{N}}{2}$ is an integer.

First, we construct the two following deterministic inputs:
\begin{itemize}

\item $\sigma^{\rm I}$: Every system observes the same measurement values all the time. There are two events ${\cal M} = \{j_1, j_2\}$ with the following setting:
\begin{equation}
w_i^{j_1} = w_i^{j_2} = \theta_i +\epsilon, \quad
{\tt t}^{j_1} = 0, \quad {\tt t}^{j_2} = 1
\end{equation}
for an arbitrarily small $\epsilon$.

\item $\sigma^{\rm II}$: A particular system is selected by the adversarial to observe non-zero measurement values, while the other systems observe only zero measurement values. There are $\frac{\sqrt{N}}{2}$ events, ${\cal M} = \{j_1, j_2, ..., j_{\frac{\sqrt{N}}{2}}\}$. Without loss of generality, we assume that system $i'$ has the minimum threshold such that $\theta_{i'} = \theta_{\min}$, which will observe non-zero measurement values as follows:
\begin{equation}
w_{i'}^{j_h} = \frac{2^{h}}{\sqrt{N}}+\epsilon, \quad {\tt t}^{j_h} = 1-\frac{1}{2^h}
\end{equation}
for an arbitrarily small $\epsilon$. And $w_i^{j_h} = 0$ for $i \ne i'$. See Fig.~\ref{fig:input1} for an illustration.

\end{itemize}

\vspace{-10pt}
\begin{figure}[htb!] 
\centering
\includegraphics[scale=0.6]{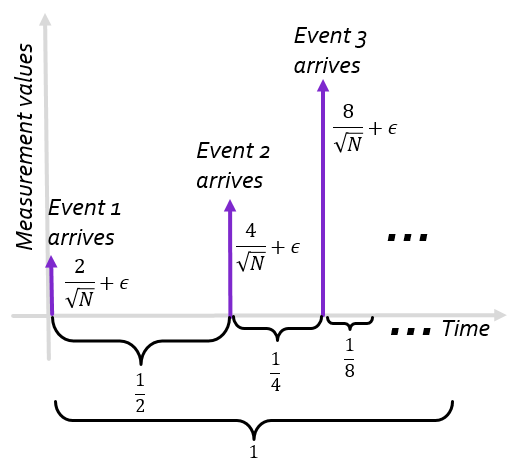}  
\vspace{-10pt}
\caption{\small   An illustration for $\sigma^{\rm II}$. \label{fig:input1}}  
%\vspace{-10pt} 
\end{figure}

If $\theta_{\min} \ge \frac{1}{\sqrt{N}}$, then the adversary chooses input $\sigma^{\rm I}$.
Otherwise, the adversary chooses input $\sigma^{\rm II}$. 

For $\sigma^{\rm I}$, ${\tt Cost}({\cal A}[\sigma^{\rm I}]) \ge \frac{1}{2}(2 + \frac{N}{ \sqrt{N}})$.
But the optimal offline cost ${\tt Opt}(\sigma^{\rm I}) \le \frac{1}{2} \cdot 2$. Therefore, ${\tt CR}({\cal A})\ge \frac{\sqrt{N}}{2} + 1$.

For $\sigma^{\rm II}$, only system $i'$ observes non-zero measurement values, which produces a report message for every event (since its threshold $\theta_{i'} < \frac{1}{\sqrt{N}}$). We obtain 
${\tt Cost}({\cal A}[\sigma^{\rm II}]) \ge \frac{1}{2} \frac{\sqrt{N}}{2}$.
But ${\tt Opt}(\sigma^{\rm II}) \le \frac{1}{2} \cdot 2$, because one report message is sufficient to be produced at the last event, incurring the total latency penalty $(\sum_{h=1}^{\frac{\sqrt{N}}{2}} \frac{2^h}{\sqrt{N}}\frac{1}{2^{h-1}}) = 1)$. Hence, ${\tt CR}({\cal A})\ge \frac{\sqrt{N}}{4}$.

{\bf Step 2}:
Next, we consider non-threshold-based algorithm ${\cal A}'$. Note that the behavior of ${\cal A}'$ is always identical after resetting $t=0$, because ${\cal A}'$ is deterministic. The optimal offline cost ${\tt Opt}(\sigma)$ does not change.  We treat ${\cal A}'$ as a blackbox. First, we input a measurement value $w_1 = \frac{2}{\sqrt{N}}$ at time $t=0$ to all systems. Suppose that there is intercommunication (possibly, a report message) produced by system $i_1$ at $t = \tau_1$. Note that any intercommunication among systems is as costly as producing a report message. If $\tau_1 \ge \frac{1}{\sqrt{N}}$, then we apply the argument using $\sigma^{\rm I}$, as in threshold-based algorithm ${\cal A}$, and obtain the competitive ratio ${\tt CR}({\cal A}')\ge \frac{\sqrt{N}}{2} + 1$.

Otherwise, $\tau_1 \le \frac{1}{\sqrt{N}}$. We input a measurement value $w_1 = \frac{2}{\sqrt{N}}$ at time $t=0$ to only system $i_1$. Since ${\cal A}'$ is deterministic, there is still intercommunication produced by system $i_1$ at least at $t = \tau_1$. Then we input a measurement value $w_2 = \frac{4}{\sqrt{N}}$ at time $t=\tau_1 + \epsilon$ to all systems. Suppose that there is intercommunication (possibly, a report message) produced by system $i_2$ at $t=\tau_1 + \epsilon + \tau_2$. If $\tau_2 \ge \frac{1}{\sqrt{N}}$, then we apply the argument using $\sigma^{\rm I}$. Note that the optimal offline cost is less than 2, because one report message is sufficient to be produced at last, incurring the total latency penalty at most $1$. We obtain the competitive ratio ${\tt CR}({\cal A}')\ge \frac{\sqrt{N}}{2} + 1$.

Iteratively, we apply the same argument up to $h$ times. If $\tau_h \le \frac{1}{\sqrt{N}}$. We input measurement values $(w_1, ..., w_{h+1})$, where $(w_1, ..., w_h)$ follows from the first $h$ measurement values of $\sigma^{\rm II}$ at systems $(i_1, ..., i_{h})$ accordingly, and we set $w_{h+1} = \frac{2^{h+1}}{\sqrt{N}}$ at time $t = \sum_{h' \le h} \tau_{h'}+ h\epsilon$ to all systems. Suppose that there is intercommunication (possibly, a report message) produced by system $i_{h+1}$ at $t=\sum_{h' \le h} \tau_{h'}+ h\epsilon+ \tau_{h+1}$. If $\tau_{h+1} \ge \frac{1}{\sqrt{N}}$, then we apply the argument using $\sigma^{\rm I}$.

Finally, if $\tau_h \le \frac{1}{\sqrt{N}}$ for all $h \le \frac{\sqrt{N}}{2}$, then we apply the argument using $\sigma^{\rm II}$, as in threshold-based algorithm ${\cal A}$, and obtain the competitive ratio ${\tt CR}({\cal A}')\ge \frac{\sqrt{N}}{4}$.
\end{IEEEproof}

\subsection{Optimal Offline Solution: A Lower Bound}

We generalize the offline algorithm in \cite{AftabChau:2013} for the 1-report problem to the setting with arbitrary non-decreasing communication cost and latency penalty functions. We denote an offline algorithm by ${\cal A}_{\rm ofl}$ which can produce a lower bound to total cost of the optimal offline solution. 
The basic idea of ${\cal A}_{\rm ofl}$ is explained as follows. First, a 2D table is created denoted by ${\tt Cost}[\cdot,\cdot]$, indexed by the set of events $\cM=\{1,...,m\}$. Each ${\tt Cost}[j, \ell]$ denotes the cost when the last report message is produced immediately after event $j$, and the next report message is produced immediately after event $\ell$. ${\tt Cost}_{\min}[j]$ is the minimum cost obtained up to event $j$.

\begin{algorithm}
\caption{\small   ${\cal A}_{\rm ofl}(K, \sigma)$} \label{alg:off}
{\small  \begin{algorithmic}[1]
\Require Number of reports per event $K$; sequence of measurements $\sigma = ({\tt t}^j, (w^j_i)_{i \in \cN})_{j\in \cM}$
%\Statex \hspace{-15pt} {\bf Init.:} 
\vspace{3pt} \hrule \vspace{3pt}
%\Statex \Comment{{\em Initialization}}
\State Let $\cM = \{1,...,m\}$
\State ${\tt Cost}[1,1] \leftarrow \rho \cdot K\cdot \min_{i\in \cN}\{\cC(\{\sigma_i^1\}) \} $ \label{alg:com1}
\State ${\tt Cost}_{\min}[1] \leftarrow {\tt Cost}[1,1]$
\For {${j} = 2,..., m$}
\State ${\tt Cost}_{\min}[j] \leftarrow \infty$
\EndFor
%\State ${\tt idx}[1] \leftarrow 1$
%\State ${\tt sys}[1] \leftarrow \bigcup_{k=1}^K\arg k^{\rm th}\mbox{-}{\tt smallest}_{i\in \cN} \{\cC(\{\sigma_i^1\})\}$
%\vspace{3pt} \hrule  \vspace{3pt}
\Statex \Comment{{\em Construct a table for dynamic programming}}
\For {${j} = 2,..., m$}
\For {${\ell} = 1,..., {j}$}
\Statex \Comment{{\em Compute latency penalty}}
\State ${\tt lat} \leftarrow \displaystyle  \sum_{ r: j-\ell < r \le j} \sum_{i \in {\cal N} } {\cal L}(\sigma_i^{r}, {\tt t}^r)$
\Statex \Comment{{\em Compute lower bound of communication cost}}
\State ${\tt com} \leftarrow K \cdot   \min_{i\in\cN}\{{\cal C}\big(\{ \sigma_i^r:  j-\ell < r \le j  \}\big) \}$ \label{alg:com2}
\\

\State ${\tt Cost}[j,\ell] \leftarrow \rho \cdot {\tt com}+ (1-\rho) \cdot {\tt lat} + {\tt Cost}_{\min}[j-\ell]$
\\

\If { ${\tt Cost}[j,\ell] < {\tt Cost}_{\min}[j]$ }
\State ${\tt Cost}_{\min}[j] \leftarrow  {\tt Cost}[j,\ell]$\label{alg:sub1}
%\State ${\tt idx}[j] \leftarrow v$ \label{alg:sub2}
%\State ${\tt sys}[j] \leftarrow$
%\Statex \qquad \qquad   $ \bigcup_{k=1}^K \arg k^{\rm th}\mbox{-}{\tt smallest}_{i\in \cN} \{{\cal C}\big(\{ \sigma_i^j:  u-v < j \le u \}\big) \} $ 
\EndIf 
\EndFor
\EndFor
%\Statex \Comment{{\em Backtracking steps of optimal solution}}
%%\State $\cR \leftarrow \cR \cup \{t_n\}$
%\State ${\cal R}_i \leftarrow \varnothing$ for all $i \in \cN$ %\{ {\tt t}^{m}\}$
%\State ${\tt sys}[m+1] \leftarrow$
%\Statex \quad  $ \bigcup_{k=1}^K \arg k^{\rm th}\mbox{-}{\tt smallest}_{i\in \cN} \{{\cal C}\big(\{ \sigma_i^j:  m-{\tt idx}[m] < j \le m \}\big) \} $ 
%\State $\cR_i \leftarrow \cR_i \cup \{ {\tt t}^m_i \}$ for all $i \in {\tt sys}[m+1]$
%
%\State $\ell \leftarrow m - {\tt idx}[m]$
%\While {$ \ell > 0$} 
%\State $\cR_i \leftarrow \cR_i \cup \{ {\tt t}^\ell_i \}$ for all $i \in {\tt sys}[\ell]$
%
%\State $\ell \leftarrow \ell - {\tt idx}[\ell]$
%
%\EndWhile
\State \Return ${\tt Cost}_{\min}[m] $
\end{algorithmic}} 
\end{algorithm}

\smallskip

\begin{customthm}{8}
Algorithm ${\cal A}_{\rm ofl}$ outputs a lower bound to the total cost of the optimal offline solution.
\end{customthm}

\smallskip

For the $1$-report problem, ${\cal A}_{\rm ofl}$ actually outputs the total cost of the optimal offline solution by dynamic programming. For the $K$-report problem, we obtain the lower bound by substituting the communication cost function by $K \cdot {\cal C}( \cdot )$. ${\cal A}_{\rm ofl}$ proceeds to compute the solution as in the $1$-report problem. This gives a solution of lower cost than the optimal offline solution, because a report message is replicated $K$ times from the same system has a lower communication cost than the one with $K$ report messages from different systems.
The rest of proof is similar to that in \cite{AftabChau:2013}. %We show that there exists a sub-optimal structure in the solution, as required by dynamic programming. Hence, ${\cal A}_{\rm ofl}$ explores every possible range $[j, \ell]$ for producing the next report message, and takes the minimum cost solution ${\tt Cost}_{\min}[j]$ from the sub-problem defined by the range $[1,j]$ in line~\ref{alg:sub1}. Note that the running time of ${\cal A}_{\rm ofl}$ is $O(N^2)$.

%This is achieved by taking the minimum cost solution for every range $[u, v]$ in line~\ref{alg:sub1}. Moreover, the offline algorithm explores all possible sub-problems and hence obtains the optimal solution.

%The proof is similar to that in \cite{AftabChau:2013}. The problem exhibits a sub-optimal structure for which the offline algorithm obtains an optimal solution for each sub-problem (by taking the minimum cost solution for every range $[u, v]$ in lines~\ref{alg:sub1}-\ref{alg:sub2}). Moreover, the offline algorithm explores all possible sub-problems and hence obtains the optimal solution.

%We mention that when the condition $w^j_i>0$ for all $i\in \cN, j\in \cM$ does not hold, the algorithm obtains a lower bound. This follows directly from the fact that the communication cost (lines \ref{alg:com1} and \ref{alg:com2}) is a lower bound on the true communication cost.

\end{document}